\documentclass[letterpaper,10pt]{IEEEtran}	
\usepackage{svn-multi}
\svnid{$Id$}

\usepackage[vlined,boxed,linesnumbered]{algorithm2e}

\usepackage{amsmath}
\usepackage{amssymb}
\usepackage{amsthm}
\usepackage{amstext}
\usepackage{tikz}
\usetikzlibrary{shapes,arrows}
\usepackage{todo}
\usepackage{alltt}
\usepackage{hyperref}
\usepackage{bbm}
\usepackage{mathrsfs}
\usepackage{lineno}
\usepackage{wrapfig}
\usepackage{mathrsfs}
\usepackage{multirow}
\usepackage{upgreek}
\usepackage{todo}
\usepackage{placeins}
\usepackage{rotating}

\DeclareMathAlphabet{\mathpzc}{OT1}{pzc}{m}{it}
\newtheorem{definition}{Definition}
\newtheorem{theorem}{Theorem}

\begin{document}

\tikzstyle{decision} = [diamond, draw,
    text width=3em, text badly centered, inner sep=0pt, aspect=2, minimum width=4.1em, minimum height=2.3em]
\tikzstyle{block} = [rectangle, draw,
    text width=3em, text centered, minimum height=2em, minimum width=4.1em]
 \tikzstyle{nobblock} = [rectangle, draw,
    text width=3em, text centered, minimum height=2em]
\tikzstyle{line} = [draw, color=black, -latex]
\tikzstyle{block2} = [rectangle, draw,
    text width=3em, text centered, rounded corners, minimum height=2em]
\tikzstyle{linewithout} = [draw, color=black]
\tikzstyle{cloud} = [draw, ellipse]

\title{Difference Constraints: An adequate Abstraction for Complexity Analysis of Imperative Programs}

\author{Moritz Sinn, Florian Zuleger, Helmut Veith\\TU Wien, Austria\thanks{Supported by the Austrian National Research
      Network S11403-N23 (RiSE) of the Austrian Science Fund (FWF) and
      by the Vienna Science and Technology Fund (WWTF) through
      grants PROSEED and ICT12-059.}
}

\maketitle

\newcommand{\loc}{\ensuremath{l}}
\newcommand{\hatloc}{\ensuremath{\hat{\loc}}}
\newcommand{\trn}{\ensuremath{\rho}}
\newcommand{\trnAlt}{\ensuremath{\tau}}
\newcommand{\edge}{\ensuremath{e}}
\newcommand{\states}{\ensuremath{\Sigma}}
\newcommand{\trns}{\ensuremath{\Gamma}}
\newcommand{\prog}{\ensuremath{\mathcal{P}}}
\newcommand{\deltaprog}{\ensuremath{\Delta\mathcal{P}}}
\newcommand{\locs}{\ensuremath{L}}
\newcommand{\edges}{\ensuremath{E}}
\newcommand{\edgesdfg}{\ensuremath{\mathcal{E}}}
\newcommand{\vars}{\ensuremath{\mathcal{V}}}
\newcommand{\symbconst}{\ensuremath{\mathcal{C}}}
\newcommand{\atoms}{\ensuremath{\mathcal{A}}}
\newcommand{\atom}{\ensuremath{\mathtt{a}}}
\newcommand{\var}{\ensuremath{\mathtt{v}}}
\newcommand{\hatvar}{\ensuremath{\mathit{\hat{\var}}}}
\newcommand{\hatatom}{\ensuremath{\mathit{\hat{\atom}}}}
\newcommand{\dcset}{\ensuremath{\mathcal{DC}}}
\newcommand{\dcp}{\ensuremath{\mathit{DCP}}} 
\newcommand{\dcps}{\ensuremath{\mathit{DCPs}}}
\newcommand{\dc}{\ensuremath{\mathit{DC}}}
\newcommand{\dcs}{\ensuremath{\mathit{DCs}}}
\newcommand{\guardset}{\ensuremath{\mathcal{G}}}
\newcommand{\guards}{\ensuremath{G}}
\newcommand{\guard}{\ensuremath{g}}
\newcommand{\update}{\ensuremath{u}}
\newcommand{\trans}{\ensuremath{\trnAlt}}
\newcommand{\transtwo}{\ensuremath{\mathtt{t}}}
\newcommand{\hattrans}{\transtwo}
\newcommand{\transitions}{\ensuremath{T}}
\newcommand{\vals}{\ensuremath{\mathit{Val}}}
\newcommand{\val}{\ensuremath{\sigma}} 
\newcommand{\trace}{\ensuremath{\rho}}
\newcommand{\edgesP}{\ensuremath{E^\prime}}
\newcommand{\norm}{\ensuremath{e}}
\newcommand{\norms}{\ensuremath{\mathit{N}}}
\newcommand{\defined}{\ensuremath{\mathcal{D}}}
\newcommand{\incr}{\ensuremath{\mathtt{c}}}
\newcommand{\hatincr}{\ensuremath{\hat{\incr}}}
\newcommand{\AllCyclic}{\ensuremath{\mathcal{I}}}
\newcommand{\AllResets}{\ensuremath{\mathcal{R}}}
\newcommand{\AllResetsPathSens}{\ensuremath{\mathcal{R}^\mathcal{L}}}
\newcommand{\AllResetsV}{\ensuremath{\mathfrak{R}}}
\newcommand{\AllResetsVSum}{\ensuremath{\mathfrak{R}_\Sigma}}
\newcommand{\ConstantV}{\ensuremath{\mathfrak{A}}}
\newcommand{\deltavar}{\ensuremath{\mathit{\delta}}}
\newcommand{\incomingvar}{\ensuremath{\mathit{in}}}
\newcommand{\id}{\ensuremath{\mathit{id}}}
\newcommand{\suffix}{\ensuremath{\mathit{suffix}}}
\newcommand{\prepost}{\ensuremath{\mathtt{PPD}}}
\newcommand{\dfg}{\ensuremath{\mathcal{G}}}

\newcommand{\LoopPathsAll}{\ensuremath{\mathcal{L}}}
\newcommand{\LoopPathsSome}{\ensuremath{\mathcal{L}_\vee}}
\newcommand{\LoopPathsSomeV}{\ensuremath{\mathfrak{L}_\vee}}
\newcommand{\paath}{\ensuremath{\pi}}
\newcommand{\paathtwo}{\ensuremath{\varpi}}
\newcommand{\ACyclic}{\ensuremath{\mathcal{A}}}
\newcommand{\LoopPathspos}{\ensuremath{\mathcal{L}^+}}
\newcommand{\LoopPathsNeg}{\ensuremath{\mathcal{L}^-}}
\newcommand{\LoopPathsposV}{\ensuremath{\mathfrak{L}^+}}
\newcommand{\ReachingDefinitions}{\ensuremath{\mathfrak{RD}}}

\newcommand{\PathBound}{\ensuremath{\mathtt{PaB}}}
\newcommand{\PotentialBound}{\ensuremath{\mathtt{PoB}}}
\newcommand{\map}{\ensuremath{\mathtt{map}}}
\newcommand{\filter}{\ensuremath{\mathtt{filter}}}
\newcommand{\reduce}{\ensuremath{\mathtt{reduce}}}
\newcommand{\VariableBound}{\ensuremath{\mathit{V\mathcal{B}}}}
\newcommand{\VariableBoundPathSens}{\ensuremath{\mathit{V\mathfrak{B}}}}
\newcommand{\expressions}{\ensuremath{\mathit{Expr}}}
\newcommand{\expr}{\ensuremath{\mathtt{expr}}}
\newcommand{\IncrCyclic}{\ensuremath{\mathtt{Incr}}}
\newcommand{\initvalue}{\ensuremath{\mathtt{init}}}
\newcommand{\Resett}{\ensuremath{\mathtt{Reset}}}
\newcommand{\reset}{\ensuremath{\mathtt{r}}}
\newcommand{\hatreset}{\ensuremath{\mathtt{\hat{r}}}}
\newcommand{\TransitionBound}{\ensuremath{\mathit{T\mathcal{B}}}}
\newcommand{\TransitionBoundPathSens}{\ensuremath{\mathit{T\mathfrak{B}}}}
\newcommand{\ProgramBound}{\ensuremath{\mathtt{\dcp Bound}}}
\newcommand{\LoopBound}{\ensuremath{\mathtt{LB}}}

\newcommand{\Decr}{\ensuremath{\mathtt{\downarrow}}}

\newcommand{\dfgpath}{\ensuremath{p}}
\newcommand{\npath}{\ensuremath{\kappa}}
\newcommand{\npathhat}{\ensuremath{\hat{\kappa}}}
\newcommand{\pattern}{\ensuremath{\mathtt{p}}}
\newcommand{\lpathstonodedfgs}{\ensuremath{\zeta}}
\newcommand{\lowerBound}{\ensuremath{\eta}}
\newcommand{\dataFlowGraph}{\ensuremath{\mathcal{D}}}

\newcommand{\potentials}{\ensuremath{\mathcal{P}}}
\newcommand{\potfun}{\ensuremath{\mathtt{p}}}

\newcommand{\PathSensIncr}{\ensuremath{\mathfrak{I}}}
\newcommand{\PathSensIncrnd}{\ensuremath{\mathscr{I}}}
\newcommand{\PathSensReset}{\ensuremath{\mathfrak{R}}}
\newcommand{\ResetsStable}{\ensuremath{\mathcal{S}_\mathfrak{R}}}
\newcommand{\ResetsUnstable}{\ensuremath{\mathcal{U}_\mathfrak{R}}}

\newcommand{\bound}{\ensuremath{\mathtt{expr}}}
\newcommand{\pot}{\ensuremath{p}}
\newcommand{\state}{\ensuremath{s}}
\newcommand{\partialorder}{\ensuremath{\succ}}

\newcommand{\relationNodesDFG}{\ensuremath{\mathfrak{R}}}
\newcommand{\numberOccTrans}{\ensuremath{\sharp}}
\newcommand{\numberOccDecr}{\ensuremath{{\downarrow}}}
\newcommand{\onefunction}{\ensuremath{\textbf{1}}}
\newcommand{\matching}{\ensuremath{m}}
\newcommand{\suffixresets}{\ensuremath{\mathcal{S}}}

\newcommand{\Loop}{\ensuremath{\mathit{L}}}
\newcommand{\LoopOnly}{\ensuremath{\mathit{L_{\cap}}}}
\newcommand{\SCC}{\ensuremath{\mathtt{SCC}}}

\newcommand{\op}{\ensuremath{\mathtt{op}}}

\newcommand{\npathin}{\ensuremath{\mathit{in}}}
\newcommand{\npathincr}{\ensuremath{\mathit{c}}}
\newcommand{\npathnodes}{\ensuremath{\mathit{atm}}}
\newcommand{\npathtrans}{\ensuremath{\mathit{trn}}}

\begin{abstract}
Difference constraints have been used for termination analysis in the literature, where they denote relational inequalities of the form $x' \le y + c$, and describe that the value of $x$ in the current state is at most the value of $y$ in the previous state plus some constant $c \in \mathbb{Z}$.
In this paper, we argue that the complexity of imperative programs typically arises from counter increments and resets, which can be modeled naturally by difference constraints.
We present the first practical algorithm for the analysis of difference constraint programs and describe how C programs can be abstracted to difference constraint programs.
Our approach contributes to the field of automated complexity and (resource) bound analysis by enabling automated amortized complexity analysis for a new class of programs and providing a conceptually simple program model that relates invariant- and bound analysis.
We demonstrate the effectiveness of our approach through a thorough experimental comparison on real world C code:
our tool Loopus computes the complexity for considerably more functions in less time than related tools from the literature.
\end{abstract}

\section{Introduction}

Automated program analysis for inferring program complexity and (resource) bounds is a very active area of research.
Amongst others, approaches have been developed for analyzing functional programs~\cite{journals/toplas/HoffmannAH12},
C\#~\cite{conf/pldi/GulwaniZ10},
C~\cite{conf/sas/AliasDFG10,conf/sas/ZulegerGSV11,sinn2014simple},
Java~\cite{journals/tcs/AlbertAGPZ12} and
Integer Transition Systems~\cite{journals/tcs/AlbertAGPZ12,conf/tacas/BrockschmidtEFFG14,conf/aplas/Flores-MontoyaH14}.

\emph{Difference constraints} ($\dcs$) have been introduced by Ben-Amram for termination analysis in~\cite{journals/toplas/Ben-Amram08}, where they denote relational inequalities of the form $x' \le y + c$, and describe that the value of $x$ in the current state is at most the value of $y$ in the previous state plus some constant $c \in \mathbb{Z}$.
We call a program whose transitions are given by a set of difference constraints a {\em difference constraint program} ($\dcp$).

In this paper, we advocate the use of $\dcs$ for program complexity and (resource) bounds analysis.
Our key insight is that $\dcs$ provide a \emph{natural abstraction} of the standard manipulations of counters in imperative programs:
counter \emph{increments/decrements} $x := x + c$ resp. \emph{resets} $x := y$, can be modeled by the $\dcs$ $x' \le x + c$ resp. $x' \le y$
(see Section~\ref{sec:abstraction} on program abstraction).
In contrast, previous approaches to bound analysis can model either only resets~\cite{conf/pldi/GulwaniZ10,conf/sas/AliasDFG10,conf/sas/ZulegerGSV11,journals/tcs/AlbertAGPZ12,conf/tacas/BrockschmidtEFFG14,conf/aplas/Flores-MontoyaH14} or increments~\cite{sinn2014simple}.
For this reason, we are able to design a more powerful analysis:
In Section~\ref{subsec:motiv_amortized} we discuss that our approach achieves \emph{amortized analysis} for a new class of programs.
In Section~\ref{subsec:inv-and-bound-analysis} we describe how our approach performs \emph{invariant analysis} by means of bound analysis.

In this paper, we establish the practical usefulness of $\dcs$ for bound (and complexity) analysis of imperative programs:
1) We propose the first algorithm for bound analysis of $\dcps$.
Our algorithm is based on the dichotomy between increments and resets.
2) We develop appropriate techniques for abstracting C programs to $\dcps$:
we describe how to extract \emph{norms} (integer-valued expressions on the program state) from C programs and how to use them as variables in $\dcps$.
We are not aware of any previous implementation of $\dcps$ for termination or bound analysis.
3) We demonstrate the effectiveness of our approach through a thorough experimental evaluation.
We present the first comparison of bound analysis tools on source code from real software projects (see Section~\ref{sec:experiments}).
Our implementation performs significantly better in time and success rate.

\section{Motivation and Related Work}
\label{sec:motivation}

\begin{figure*}
 \begin{tabular}{l|l|l|l}
 \begin{minipage}{3.5cm}
 \center\small
 \begin{alltt}
void foo(uint n) \{
   int x = n;
   int r = 0;
\(l\sb{1}\)  while(x > 0) \{
     x = x - 1;
     r = r + 1;
\(l\sb{2}\)    if(*) \{
       int p = r;
\(l\sb{3}\)      while(p > 0)
         p--;
       r = 0;
     \}
\(l\sb{4}\)  \} \}
 \end{alltt}
 \vspace{-0.2cm}
 \end{minipage}
&
\hspace{-0.5cm}
\begin{minipage}{5.5cm}
  \begin{tikzpicture}[scale=0.4, node distance = 2cm, auto]

\node (t0)  {$\loc_b$};
\node (t1) [below of=t0, node distance = 1cm] {$\loc_1$};
\node (te) [right of=t1, node distance = 1cm] {$\loc_e$};
\node (t2) [below of=t1, node distance = 1cm]  {$\loc_2$};
\node (t4) [below of=t2, right of=t2, node distance = 1.5cm]  {$\loc_3$};
\node (t5) [below of=t2, left of=t2, node distance = 1.5cm]  {$\loc_4$};

\path
(t0) edge [line] node [right,font=\scriptsize ] {
 $\trans_0 \equiv $\begin{tabular}{l}
 $x^\prime \le n$;\\
 $r^\prime \le 0$;\\
 \end{tabular}
}(t1)
(t1) edge [line] node [right,font=\scriptsize ] {
\begin{tabular}{c}
$\trans_1 \equiv$
\begin{tabular}{l}
$x > 0,$\\
$x^\prime \le x - 1$\\
$r^\prime \le r + 1$\\
\end{tabular}
\end{tabular}
}(t2)
(t2) edge [line] node [right,font=\scriptsize ] {
$\trans_{2a} \equiv $\begin{tabular}{l}
$x^\prime \le x$\\
$r^\prime \le r$\\
$p^\prime \le r$\\
\end{tabular}
}(t4)
(t2) edge [line] node [right,near end,font=\scriptsize,yshift=-1pt,xshift=-2pt] {
$\trans_{2b}\equiv$\begin{tabular}{l}
$x^\prime \le x$\\
$r^\prime \le r$\\
\end{tabular}}
(t5)
(t4) edge [line] node [below,font=\scriptsize ] {
$\trans_4 \equiv $\begin{tabular}{l}
$x^\prime \le x$\\
$r^\prime \le 0$\\
\end{tabular}
}(t5)
(t5) edge [line, bend left] node [left,font=\scriptsize ] {
\begin{tabular}{l}
$\trans_5 \equiv $\\
$r^\prime \le r$\\
$x^\prime \le x$\\
\end{tabular}}(t1)
(t4) edge [loop below] node [right,font=\scriptsize] {
\begin{tabular}{l}
$p > 0,$\\
$x^\prime \le x$\\
$r^\prime \le r$\\
$p^\prime \le p - 1$\\
\end{tabular}
} node [below,font=\scriptsize] {
\begin{tabular}{c}
$\trans_3 \equiv$
\end{tabular}
}
(t4)
(t1) edge [line] (te)
;

\end{tikzpicture}
\end{minipage} &

 \begin{minipage}{3.5cm}
  \center\small
  \begin{alltt}
foo(uint n, uint m1,
    uint m2) \{
  int y = n;
  int x;
\(l\sb{1}\) if(*)
     x = m1;
   else
     x = m2;
\(l\sb{2}\)  while(y > 0) \{
     y--;
     x = x + 2; \}
   int z = x;
\(l\sb{3}\)  while(z > 0)
     z--; \}
  \end{alltt}
  \vspace{-0.2cm}
  \end{minipage}

  &
  \hspace{-0.5cm}
\begin{minipage}{3cm}
  \center
  \begin{tikzpicture}[scale=0.4, node distance = 2cm, auto]

\node (t0)  {$\loc_b$};
\node (t00) [below of=t0, node distance = 1cm] {$\loc_1$};
\node (t0b) [below of=t00, node distance = 1cm] {$\loc_2$};
\node (t1) [below of=t0b, node distance = 1cm] {$\loc_3$};
\node (te) [right of=t1, node distance = 1cm] {$\loc_e$};

\path
(t0) edge [line] node [right,font=\scriptsize ] {
 $\trans_{0} \equiv $\begin{tabular}{l}
 $y^\prime \le n$\\
 \end{tabular}
}(t00)
(t00) edge [line] node [right,font=\scriptsize,at start] {
 $\trans_{0a} \equiv $\begin{tabular}{l}
 $y^\prime \le y$\\
 $x^\prime \le m1$\\
 \end{tabular}
}(t0b)
(t00) edge [line,bend right] node [left,font=\scriptsize ] {
 \begin{tabular}{l}
 $\trans_{0b} \equiv $\\
 $y^\prime \le y$\\
 $x^\prime \le m2$\\
 \end{tabular}
}(t0b)
(t0b) edge [loop right] node [right,font=\scriptsize] {
\begin{tabular}{l}
$\trans_1 \equiv$
\begin{tabular}{l}
$y > 0,$\\
$y^\prime \le y - 1$\\
$x^\prime \le x + 2$\\
\end{tabular}
\end{tabular}}(t0b)
(t0b) edge [line] node [right,font=\scriptsize ] {
 $\trans_2 \equiv $\begin{tabular}{l}
 $z^\prime \le x$;
 \end{tabular}
}(t1)
(t1) edge [loop below] node [below,font=\scriptsize] {
$\trans_3 \equiv$
\begin{tabular}{c}
$z > 0,$\\
$z^\prime \le z - 1$\\
\end{tabular}}(t1)
(t1) edge [line] (te)
;

\end{tikzpicture}
\end{minipage} \\
\multicolumn{2}{c|}{\scriptsize Complexity: $\TransitionBound(\trans_5) + \TransitionBound(\trans_3) = n + n = 2n$} & \multicolumn{2}{c}{\scriptsize Complexity: $\TransitionBound(\trans_1) + \TransitionBound(\trans_3) = \max(m_1,m_2) + 3n$}\\
\scriptsize Example 1 & \scriptsize  abstracted \dcp\ of Example 1 &  \scriptsize  Example 2 &\scriptsize  abstracted \dcp\ of Example 2\\
\end{tabular}
\caption{Running Examples, * denotes non-determinism (arising from conditions not modeled in the analysis)}
\label{fig:ex1}
\end{figure*}
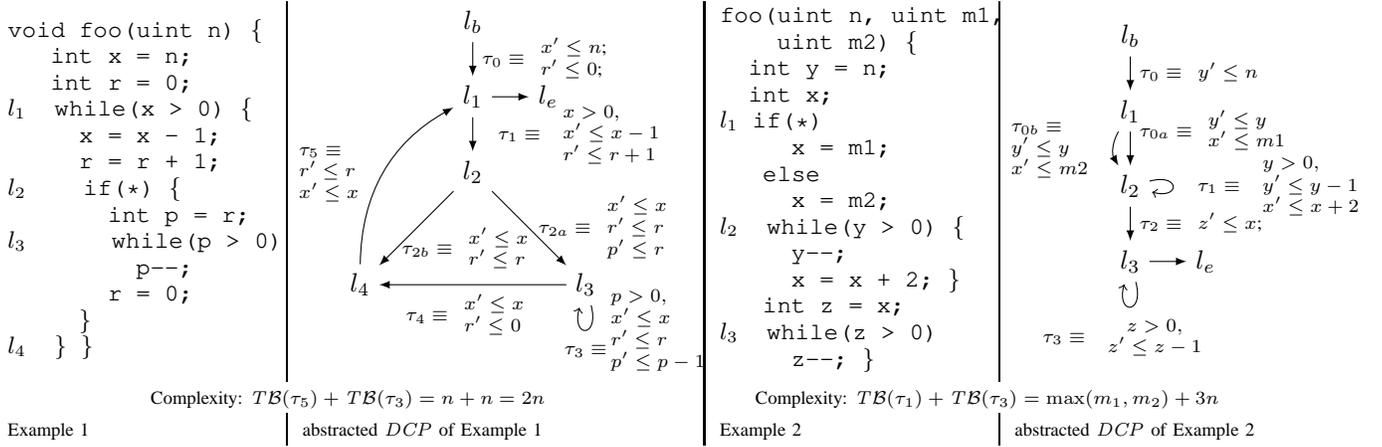

\subsection{Amortized Complexity Analysis}
\label{subsec:motiv_amortized}

Example~1 stated in Figure~\ref{fig:ex1} is representative for a class of loops that we found in parsing and string matching routines during our experiments.
In these loops the inner loop iterates over disjoint partitions of an array or string, where the partition sizes are determined by the program logic of the outer loop.
For an illustration of this iteration scheme, we refer the reader to Example~3 stated {in Appendix~\ref{app-example}},
which contains a snippet of the source code after which we have modeled Example~1.
Example~1 has the linear \emph{complexity} $2n$, because the inner loop as well as the outer loop can be iterated at most $n$ times (as argued in the next paragraph).
However, previous approaches to bound analysis~\cite{conf/pldi/GulwaniZ10,conf/sas/AliasDFG10,conf/sas/ZulegerGSV11,sinn2014simple,journals/tcs/AlbertAGPZ12,conf/tacas/BrockschmidtEFFG14,conf/aplas/Flores-MontoyaH14} are only able to deduce that the inner loop can be iterated at most a \emph{quadratic} number of times (with loop bound $n^2$) by the following reasoning:
(1) the outer loop can be iterated at most $n$ times,
(2) the inner loop can be iterated at most $n$ times \emph{within} one iteration of the outer loop (because the inner loop has a local loop bound $p$ and $p \le n$ is an invariant),
(3) the loop bound $n^2$ is obtained from (1) and (2) by multiplication.
We note that inferring the linear complexity $2n$ for Example~1, even though the inner loop can already be iterated $n$ times \emph{within} one iteration of the outer loop, is an instance of {\em amortized complexity analysis}~\cite{amortizedComplexity}.

In the following, we give an overview how our approach infers the linear complexity for Example~1:\\
\textbf{1. Program Abstraction.}
We abstract the program to a \dcp\ over $\mathbb{Z}$ as shown in Figure~\ref{fig:ex1}. 
We discuss our algorithm for abstracting imperative programs to \dcp s based on symbolic execution  in Section~\ref{sec:abstraction}.\\
\textbf{2. Finding Local Bounds.}
We identify $p$ as a variable that limits the number of executions of transition $\trans_3$:
We have the guard $p > 0$ on $\trans_3$ and $p$ decreases on each execution of $\trans_3$.
We call $p$ a \emph{local bound} for $\trans_3$.
Accordingly we identify $x$ as a {\em local bound} for transitions $\trans_1,\trans_{2a},\trans_{2b},\trans_{4},\trans_{5}$.\\
\textbf{3. Bound Analysis.}
Our algorithm (stated in Section~\ref{sec-alg}) computes {\em transition bounds}, i.e., (symbolic) upper bounds on the number of times program transitions can be executed, and {\em variable bounds}, i.e., (symbolic) upper bounds on variable values.
For both types of bounds, the main idea of our algorithm is to reason {\em how much} and {\em how often} the value of the local bound resp. the variable value may increase during program run.
Our algorithm is based on a mutual recursion between variable bound analysis (``how much'', function $\VariableBound(\var)$) and transition bound analysis (``how often'', function $\TransitionBound(\trans)$).
Next, we give an intuition how our algorithm computes transition bounds:
Our algorithm computes $\TransitionBound(\trans) = n$ for $\trans \in \{\trans_1,\trans_{2a},\trans_{2b},\trans_4,\trans_5\}$ because the local bound $x$ is initially set to $n$ and never increased or reset.
Our algorithm computes $\TransitionBound(\trans_3)$ ($\trans_3$ corresponds to the loop at $\loc_3$) as follows:
$\trans_3$ has local bound $p$;
$p$ is reset to $r$ on $\trans_{2a}$;
our algorithm detects that before each execution of $\trans_{2a}$, $r$ is reset to $0$ on either $\trans_0$ or $\trans_4$, which we call the {\em context} under which $\trans_{2a}$ is executed;
our algorithm establishes that between being reset and flowing into $p$ the value of $r$ can be incremented up to $\TransitionBound(\trans_1)$ times by $1$;
our algorithm obtains $\TransitionBound(\trans_1) = n$ by a recursive call;
finally, our algorithm calculates $\TransitionBound(\trans_3) = 0 + \TransitionBound(\trans_1) \times 1 = n$.
We give an example for the mutual recursion between $\TransitionBound$ and $\VariableBound$ in Section~\ref{subsec:inv-and-bound-analysis}.

{
We contrast our approach for computing the loop bound of $\loc_3$ of Example~1 with classical invariant analysis:
Assume '$c$' counting the number of inner loop iterations (i.e., $c$ is initialized to 0 and incremented in the inner loop).
For inferring $c<=n$ through invariant analysis the invariant $c+x+r<=n$ is needed for the outer loop, and the invariant $c+x+p<=n$ for the inner loop. 
Both relate 3 variables and cannot be expressed as (parametrized) octagons (e.g.,~\cite{gawlitza2014parametric}).
Further, the expressions $c+x+r$ and $c+x+p$ do not appear in the program, which is challenging for template based approaches to invariant analysis.
}

\subsection{Invariants and Bound Analysis}
\label{subsec:inv-and-bound-analysis}

We explain on Example~2 in Figure~\ref{fig:ex1} how our approach performs \emph{invariant analysis} by means of bound analysis.
We first motivate the importance of invariant analysis for bound analysis.
It is easy to infer $x$ as a bound for the possible number of iterations of the loop at $\loc_3$.
However, in order to obtain a bound in the \emph{function parameters} the difficulty lies in finding an invariant $x \le \expr(n,{m_1},{m_2})$.
{
Here, the most precise invariant $x \le \max(m_1,m_2) + 2n$ cannot be computed by standard abstract domains such as \emph{octagon} or \emph{polyhedra}:
these domains are {\em convex} and cannot express non-convex relations such as \emph{maximum}.}
 The most precise approximation of $x$ in the polyhedra domain is $x \le m_1 + m_2 + 2n$.
Unfortunately, it is well-known that the polyhedra abstract domain does not scale to larger programs and needs to rely on heuristics for termination.
Next, we explain how our approach computes invariants using bound analysis and discuss how our reasoning is substantially different from invariant analysis by abstract interpretation.

Our algorithm computes a transition bound for the loop at $\loc_3$ by
$\TransitionBound(\trans_3) = \TransitionBound(\trans_2) \times \VariableBound(x) = 1 \times \VariableBound(x) = \VariableBound(x) = \TransitionBound(\trans_1) \times 2 + \max(m_1,m_2) = (n \times \TransitionBound(\trans_0)) \times 2 + \max(m_1,m_2) = (n \times 1) \times 2 + \max(m_1,m_2) = 2n + \max(m_1,m_2)$.
We point out the mutual recursion between $\TransitionBound$ and $\VariableBound$:
$\TransitionBound(\trans_3)$ has called $\VariableBound(x)$, which in turn called $\TransitionBound(\trans_1)$.
We highlight that the variable bound $\VariableBound(x)$ (corresponding to the invariant $x \le \max(m_1,m_2) + 2n$) has been established during the computation of $\TransitionBound(\trans_3)$.

Standard \emph{abstract domains} such as \emph{octagon} or \emph{polyhedra} propagate information \emph{forward} until a fixed point is reached, \emph{greedily} computing all possible invariants expressible in the abstract domain at every location of the program.
In contrast, $\VariableBound(x)$ infers the invariant $x \le \max(m1,m2) + 2n$ by modular reasoning:
{\em local information} about the program (i.e., increments/resets of variables, local bounds of transitions) is combined to a {\em global} program property.
Moreover, our variable and transition bound analysis is \emph{demand-driven}:
our algorithm performs only those recursive calls that are indeed needed to derive the desired bound.
We believe that our analysis complements existing techniques for invariant analysis and will find applications outside of bound analysis.

\subsection{Related Work}
In~\cite{journals/toplas/Ben-Amram08} it is shown that termination of $\dcps$ is undecidable in general but decidable for the natural syntactic subclass of {\em deterministic} $\dcps$ (see Definition~\ref{def:dcp}), which is the class of $\dcps$ we use in this paper.
It is an open question for future work whether there is a complete algorithm for bound analysis of deterministic $\dcps$.

In~\cite{sinn2014simple} a bound analysis based on constraints of the form $x' \le x + c$ is proposed, where $c$ is either an integer or a symbolic constant.
The resulting abstract program model is strictly less powerful than $\dcps$.
In~\cite{conf/sas/ZulegerGSV11} a bound analysis based on so-called \emph{size-change constraints} $x' \lhd y$ is proposed, where $\lhd \in \{<,\le\}$.
Size-change constraints form a strict syntactic subclass of $\dcs$.
However, termination is decidable even for non-deterministic size-change programs and a complete algorithm for deciding the complexity of size-change programs has been developed~\cite{conf/mfcs/ColcombetDZ14}.
Because the constraints in~\cite{conf/sas/ZulegerGSV11,sinn2014simple} are less expressive than $\dcs$, the resulting bound analyses cannot infer the linear complexity of Example~1 and need to rely on external techniques for invariant analysis.

In Section~\ref{sec:experiments} we compare our implementation against the most recent approaches to automated complexity analysis~\cite{conf/aplas/Flores-MontoyaH14,conf/tacas/BrockschmidtEFFG14,sinn2014simple}.
\cite{conf/aplas/Flores-MontoyaH14} extends the COSTA approach by control flow refinement for cost equations and a better support for multi-dimensional ranking functions.
The COSTA project (e.g.~\cite{journals/tcs/AlbertAGPZ12}) computes resource bounds by inferring an upper bound on the solutions of certain recurrence equations (so-called \emph{cost equations}) relying on external techniques for invariant analysis (which are not explicitly discussed).
The bound analysis in~\cite{conf/tacas/BrockschmidtEFFG14} uses approaches for computing polynomial ranking functions from the literature to derive bounds for SCCs in isolation and then expresses these bounds in terms of the function parameters using invariant analysis (see next paragraph).

The powerful idea of expressing locally computed loop bounds in terms of the function parameters by alternating between loop bound analysis and variable upper bound analysis has been explored in \cite{conf/tacas/BrockschmidtEFFG14}, \cite{sinn2014simple} (as discussed in the extended version~\cite{journals/corr/SinnZV14}) and \cite{backward_symb_exec}.
We highlight some important differences to these earlier works.
\cite{conf/tacas/BrockschmidtEFFG14} computes upper bound invariants only for the \emph{absolute} values of variables; this does, for example, not allow to distinguish between variable increments and decrements during the analysis.
\cite{journals/corr/SinnZV14} and \cite{backward_symb_exec} do not give a general algorithm but deal with specific cases.

{\cite{wu2003induction}~discusses automatic parallelization of loop iterations;
the approach builds on summarizing inner loops by multiplying the increment of a variable on a single iteration of a loop with the loop bound.
The loop bounds in~\cite{wu2003induction} are restricted to simple syntactic patterns.}

{
The recent paper \cite{carbonneaux2015compositional} discusses an interesting alternative for amortized complexity analysis of imperative programs:
A system of linear inequalities is derived using Hoare-style proof-rules.
Solutions to the system represent valid \emph{linear} resource bounds.
Interestingly, \cite{carbonneaux2015compositional} is able to compute the linear bound for $\loc_3$ of Example~1 but fails to deduce the bound for the original source code (provided in Appendix~\ref{app-example}).
Moreover, \cite{carbonneaux2015compositional} is restricted to linear bounds, while our approach derives polynomial bounds (e.g., Example~B in Figure~\ref{fig-ex-dcps}) which may also involve the maximum operator.
An experimental comparison was not possible as \cite{carbonneaux2015compositional} was developed in parallel. 
}
\section{Program Model and Algorithm}
\label{sec-alg}

In this section we present our algorithm for computing worst-case upper bounds on the number of executions of a given transition (transition bound) and on the value of a given variable (variable bound).
We base our algorithm on the abstract program model of \dcp s stated in Definition~\ref{def:dcp}.
In Section~\ref{subsec:alg-nonwellfounded} we generalize \dcp s and our algorithm to the non-well-founded domain $\mathbb{Z}$.

\begin{definition}[Variables, Symbolic Constants, Atoms]
 By $\vars$ we denote a finite set of Variables. By $\symbconst$ we denote a finite set of symbolic constants. $\atoms = \vars \cup \symbconst \cup \mathbb{N}$ is the set of {\em atoms}.
\end{definition}

\begin{definition}[Difference Constraints]
  A \emph{difference constraint} over $\atoms$ is an inequality of form $x^\prime \le y + c$ with $x \in \vars$, $y \in \atoms$ and $c \in \mathbb{Z}$.
  We denote by $\dcset(\atoms)$ the set of all difference constraints over $\atoms$.
\end{definition}

\begin{definition}[Difference Constraint Program]
\label{def:dcp}
  A \emph{difference constraint program} (\dcp) over $\atoms$ is a directed labeled graph $\deltaprog = (\locs, \transitions, \loc_b,\loc_e)$, where $\locs$ is a finite set of \emph{locations}, $\loc_b \in \locs$ is the entry location, $\loc_e \in \locs$ is the exit location and $\transitions \subseteq \locs \times 2^{\dcset(\atoms)} \times \locs$ is a finite set of transitions.
  We write $\loc_1 \xrightarrow{\update} \loc_2$ to denote a transition $(\loc_1, \update, \loc_2) \in \transitions$ labeled by a set of difference constraints $\update \in 2^{\dcset(\atoms)}$.
  Given a transition $\trans = \loc_1 \xrightarrow{\update} \loc_2 \in \transitions$ of $\deltaprog$ we call $\loc_1$ the source location of $\trans$ and $\loc_2$ the target location of $\trans$.
  A \emph{path} of $\deltaprog$ is a sequence $\loc_0 \xrightarrow{\update_0} \loc_1 \xrightarrow{\update_1} \cdots$ with $\loc_i \xrightarrow{\update_i} \loc_{i+1} \in \transitions$ for all $i$.
  The set of \emph{valuations} of $\atoms$ is the set $\vals_\atoms = \atoms \rightarrow \mathbb{N}$ of mappings from $\atoms$ to the natural numbers with $\val(\atom) = \atom$ if $\atom \in \mathbb{N}$.
  A \emph{run} of $\deltaprog$ is a sequence $(\loc_b,\val_0) \xrightarrow{\update_0} (\loc_1,\val_1) \xrightarrow{\update_1} \cdots$ such that $\loc_b \xrightarrow{\update_0} \loc_1 \xrightarrow{\update_1} \cdots $ is a path of $\deltaprog$ and for all $i$ it holds that (1) $\val_i \in \vals_\atoms$, (2) $\val_{i+1}(x) \le \val_i(y) + c$ for all ${x^\prime \le y + c} \in \update_i$, (3) $\val_i(s) = \val_0(s)$ for all $s \in \symbconst$.
  {Given $\var \in \vars$ and $\loc \in \locs$ we say that $\var$ is \emph{defined} at $\loc$ and write $\var \in \defined(\loc)$ if $\loc \ne \loc_b$ and for all incoming transitions ${\loc_1 \xrightarrow{\update} \loc} \in \transitions$ of $\loc$ it holds that there are $\atom \in \atoms$ and $\incr \in \mathbb{Z}$ s.t. $\var^\prime \le {\atom + \incr} \in \update$.}
  
  {\deltaprog\ is \emph{deterministic} ({\em fan-in-free} in the terminology of \cite{journals/toplas/Ben-Amram08}), if for every transition $\loc_1 \xrightarrow{\update} \loc_2 \in \transitions$ and every $\var \in \vars$ there is at most one $\atom \in \atoms$ and $\incr \in \mathbb{Z}$ s.t. $\var^\prime \le \atom + \incr \in \update$.}
\end{definition}


Our approach assumes the given \dcp\ to be {\em deterministic}.
{We further assume that \dcp s are {\em well-defined}: Let $\var \in \vars$ and $\loc \in \locs$, if $\var$ is \emph{live} at $\loc$ then $\var \in \defined(\loc)$.} Our abstraction algorithm from Section~\ref{sec:abstraction} generates only deterministic and well-defined \dcp s.

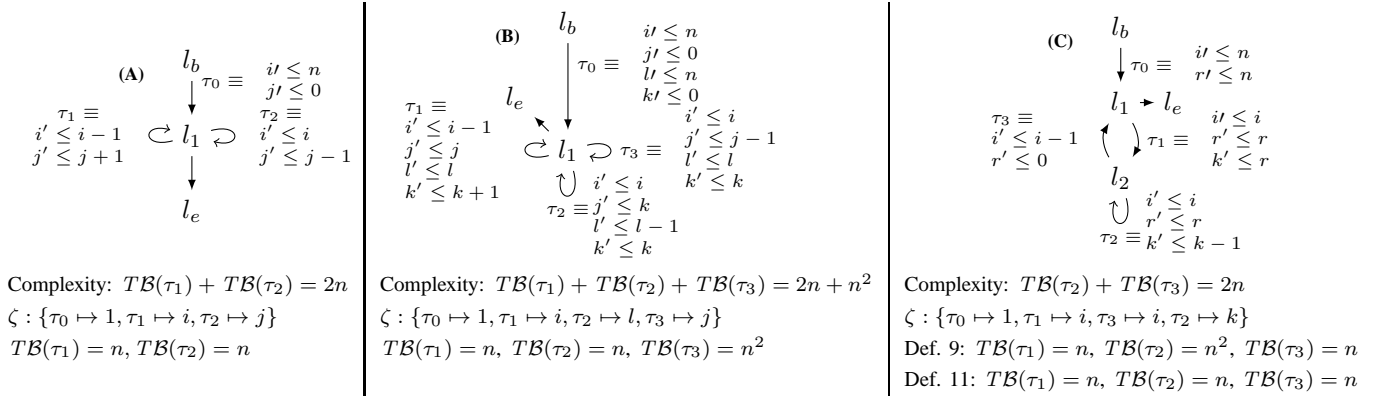
\begin{figure*}

 \begin{tabular}{l|l|l}
 \begin{minipage}{4.5cm}
 \center
   \begin{tikzpicture}[scale=0.4, node distance = 2cm, auto]
    \node (name) at (-2,-0.5) [font=\scriptsize]{\textbf{(A)}};
    \node (t0)  {$\loc_b$};
    \node (t1) [below of=t0, node distance = 1cm] {$\loc_1$};
    \node (texit) [below of=t1, node distance = 1cm] {$\loc_e$};
    \path(t0) edge [line] node [right,font=\scriptsize, at start] {
    $\trans_0 \equiv$
    \begin{tabular}{l}
    $i\prime \le n$\\
    $j\prime \le 0$
    \end{tabular}
    }(t1)
    (t1) edge [loop left] node [left,font=\scriptsize] {
    \begin{tabular}{c}
    $\trans_1 \equiv$\\
    $i^\prime \le i - 1$\\
    $j^\prime \le j + 1$
    \end{tabular}}(t1)
    (t1) edge [loop right] node [right,font=\scriptsize] {
    \begin{tabular}{l}
    $\trans_2 \equiv$\\
    $i^\prime \le i$\\
    $j^\prime \le j - 1$
    \end{tabular}}(t1)
    (t1) edge [line] (texit)
    ;
   \end{tikzpicture}
 \end{minipage}
 &
\begin{minipage}{5.5cm}
\center
   \begin{tikzpicture}[scale=0.4, node distance = 2cm, auto]
   \node (name) at (-2,-0.5) [font=\scriptsize]{\textbf{(B)}};
    \node (t0)  {$\loc_b$};
    \node (t1) [below of=t0, node distance = 1.7cm] {$\loc_1$};
    \node (texit1) [above of=t1, node distance = 0.7cm] {};
    \node (texit) [left of=texit1, node distance = 0.7cm] {$\loc_e$};
    \path(t0) edge [line] node [right,font=\scriptsize, near start] {
    $\trans_0 \equiv$
    \begin{tabular}{c}
    $i\prime \le n$\\
    $j\prime \le 0$\\
    $l\prime \le n$\\
    $k\prime \le 0$\\
    \end{tabular}
    }(t1)
    (t1) edge [loop left] node [left,font=\scriptsize] {
    \begin{tabular}{l}
    $\trans_1 \equiv$\\
    $i^\prime \le i - 1$\\
    $j^\prime \le j$\\
    $l^\prime \le l$\\
    $k^\prime \le k + 1$\\
    \end{tabular}}(t1)
    (t1) edge [loop right] node [right,font=\scriptsize] {
    $\trans_3 \equiv$
    \begin{tabular}{l}
    $i^\prime \le i$\\
    $j^\prime \le j - 1$\\
    $l^\prime \le l$\\
    $k^\prime \le k$\\
    \end{tabular}}(t1)
    (t1) edge [loop below] node [right,font=\scriptsize] {
    \begin{tabular}{l}
    \\
    \\
    $i^\prime \le i$\\
    $j^\prime \le k$\\
    $l^\prime \le l - 1$\\
    $k^\prime \le k$\\
    \end{tabular}}
    node [below,font=\scriptsize] {
    $\trans_2 \equiv$
    }
    (t1)
    (t1) edge [line] (texit)
    ;
   \end{tikzpicture}
 \end{minipage}
 &
 \begin{minipage}{6cm}
  \center
   \begin{tikzpicture}[scale=0.4, node distance = 2cm, auto]
    \node (name) at (-2,-0.5) [font=\scriptsize]{\textbf{(C)}};
    \node (t0)  {$\loc_b$};
    \node (t1) [below of=t0, node distance = 1cm] {$\loc_1$};
    \node (texit) [right of=t1, node distance = 0.7cm] {$\loc_e$};
    \node (t2) [below of=t1, node distance = 1cm] {$\loc_2$};
    \path(t0) edge [line] node [right,font=\scriptsize ] {
    $\trans_0 \equiv$
    \begin{tabular}{l}
    $i\prime \le n$\\
    $r\prime \le n$
    \end{tabular}
    }(t1)
    (t1) edge [line,bend left] node [right,font=\scriptsize] {
    $\trans_1 \equiv$
     \begin{tabular}{l}
     $i\prime \le i$\\
     $r^\prime \le r$\\
     $k^\prime \le r$\\
     \end{tabular}
    }(t2)
    (t2) edge [loop below] node [right,font=\scriptsize] {
    \begin{tabular}{l}
    $i^\prime \le i$\\
    $r^\prime \le r$\\
    $k^\prime \le k - 1$\\
    \end{tabular}} node [below,font=\scriptsize]{$\trans_2 \equiv$}(t2)
    (t2) edge [line,bend left] node [left,font=\scriptsize] {
    \begin{tabular}{l}
    $\trans_3 \equiv$\\
    $i^\prime \le i - 1$\\
    $r^\prime \le 0$
    \end{tabular}
    } (t1)
    (t1) edge [line] (texit)
    ;
   \end{tikzpicture}
 \end{minipage}
\\
 \footnotesize Complexity: $\TransitionBound(\trans_1) + \TransitionBound(\trans_2) = 2n$ & \footnotesize Complexity: $\TransitionBound(\trans_1) + \TransitionBound(\trans_2) + \TransitionBound(\trans_3) = 2n + n^2$ & \footnotesize Complexity: $\TransitionBound(\trans_2) + \TransitionBound(\trans_3) = 2n$\\
 \footnotesize $\lpathstonodedfgs: \{\trans_0 \mapsto 1, \trans_{1} \mapsto i, \trans_2 \mapsto j\}$ & \footnotesize $\lpathstonodedfgs: \{\trans_0 \mapsto 1, \trans_{1} \mapsto i, \trans_2 \mapsto l, \trans_{3} \mapsto j\}$ & \footnotesize $\lpathstonodedfgs: \{\trans_0 \mapsto 1, \trans_1 \mapsto i, \trans_3 \mapsto i, \trans_2 \mapsto k\}$\\
 \footnotesize $\TransitionBound(\trans_{1}) = n, \TransitionBound(\trans_{2}) = n$ & \footnotesize $\TransitionBound(\trans_{1}) = n$, $\TransitionBound(\trans_{2}) = n$, $\TransitionBound(\trans_{3}) = n^2$ & \footnotesize Def.~\ref{def:functions01}: $\TransitionBound(\trans_1) = n$, $\TransitionBound(\trans_2) = n^2$, $\TransitionBound(\trans_3) = n$\\
 & & \footnotesize Def.~\ref{def:functions02}: $\TransitionBound(\trans_1) = n$, $\TransitionBound(\trans_2) = n$, $\TransitionBound(\trans_3) = n$\\
 \end{tabular}
 \caption{Example \dcp's (A), (B), (C)}
 \label{fig-ex-dcps}
\end{figure*}


{In Definitions~\ref{def:transbound} to~\ref{def:functions02} we assume a \dcp\ $\deltaprog(\locs, \transitions, \loc_b, \loc_e)$ over $\atoms$ to be given.}

\begin{definition}[Transition Bound]\label{def:transbound}
{
Let $\trans \in \transitions$, $\trans$ is \emph{bounded} iff $\trans$ appears a finite number of times on any run of $\deltaprog$.
An expression $\bound$ over $\symbconst \cup \mathbb{Z}$ is a {\em transition bound} for $\trans$ iff $\trans$ is \emph{bounded} and} for any {\em finite} run $\trace = (\loc_{b}, \val_0) \xrightarrow{\update_0} (\loc_{1}, \val_1) \xrightarrow{\update_1} (\loc_{2}, \val_2) \xrightarrow{\update_2} \dots (\loc_{e}, \val_n)$ of $\deltaprog$ it holds that $\trans$ appears not more than $\val_0(\bound)$ often on $\trace$.
We say that a transition bound $\bound$ of $\trans$ is {\em precise} iff there is a run $\trace$ of $\deltaprog$ s.t. $\trans$ appears $\val_0(\bound)$ times on $\trace$.
\end{definition}

We want to infer the complexity of the examples in Figure~\ref{fig-ex-dcps} (Examples~A, B, C), i.e., we want to infer how often location $\loc_1$ can be visited during an execution of the program. 
We will do so by computing a bound on the number of times transitions $\trans_0$, $\trans_1$, $\trans_2$ and $\trans_3$ may be executed.
In general, the complexity of a given program can be inferred by summing up the transition bounds for the back edges in the program.

\begin{definition}[Counter Notation]\label{def:counternotation}
Let $\trans \in \transitions$ and $\var \in \vars$.
Let $\trace = (\loc_b,\val_0) \xrightarrow{\update_0} (\loc_1,\val_1) \xrightarrow{\update_1} \cdots (\loc_e,\val_n)$ be a finite run of $\deltaprog$.
By $\numberOccTrans(\trans, \trace)$ we denote the number of times that $\trans$ occurs on $\trace$. By $\numberOccDecr(\var, \trace)$ we denote the number of times that the value of $\var$ decreases on $\trace$, i.e.
 $\numberOccDecr(\var,\trace) = |\{i \mid {\val_i(\var) > \val_{i+1}(\var) }\}|$.
\end{definition}

\begin{definition}[Local Transition Bound]\label{def:potfun}
Let $\trans \in \transitions$ and $\var \in \vars$.
$\var$ is a {\em local bound} for $\trans$  iff on all finite runs $\trace = (\loc_b,\val_0) \xrightarrow{\update_0} (\loc_1,\val_1) \xrightarrow{\update_1} \cdots (\loc_e,\val_n)$ of $\deltaprog$ it holds that
$\numberOccTrans(\trans, \trace) \le \numberOccDecr(\var, \trace)$.

We call a {\em complete} mapping $\lpathstonodedfgs: \transitions \rightarrow \vars \cup \{\onefunction\}$ a {\em local bound mapping} for $\deltaprog$ if $\lpathstonodedfgs(\trans)$ is a {\em local bound} of $\trans$ or $\lpathstonodedfgs(\trans) = \onefunction$ and $\trans$ can only appear at most once on any path of $\deltaprog$.
\end{definition}
{\em Example A:} $i$ is a local bound for $\trans_1$, $j$ is a local bound for $\trans_2$. {\em Example C:} $i$ is a local bound for $\trans_1$ and for $\trans_3$.\\

A variable $\var$ is a {\em local transition bound} if on any run of $\deltaprog$ we can traverse $\trans$ not more often than the number of times the value of $\var$ decreases.
I.e., a local bound $\var$ limits the potential number of executions of $\trans$ as long as the value of $\var$ does not increase.
In our analysis, {\em local transition bounds} play the role of {\em potential functions} in classical {\em amortized complexity analysis}~\cite{amortizedComplexity}.
Our bound algorithm is based on a mapping which assigns each transition a local bound. We discuss how we find local bounds in Section~\ref{subsec:findpotentials}.

%

\begin{definition}[Variable Bound]
An expression $\bound$ over $\symbconst \cup \mathbb{Z}$ is a {\em variable bound} for $\var \in \vars$ iff for any finite run $\trace = (\loc_{b}, \val_0) \xrightarrow{\update_0} (\loc_{1}, \val_1) \xrightarrow{\update_1} (\loc_{2}, \val_2) \xrightarrow{\update_2} \dots (\loc_{e}, \val_n)$ of $\deltaprog$ and all $1  \le i \le n$ with $\var \in \defined(\loc_i)$ it holds that $\val_i(\var) \le \val_0(\bound)$.
\label{def:variablebound}
\end{definition}

Let $\var \in \vars$.
Our algorithm is based on a {\em syntactic} distinction between transitions which {\em increment} $\var$  or {\em reset} $\var$.

\begin{definition}[Resets and Increments]
  Let $\var \in \vars$.
  We define the \emph{resets} $\AllResets(\var)$ and \emph{increments} $\AllCyclic(\var)$ of $\var$ as follows:\\
  \begin{math}
  \begin{array}{ll}
    \AllResets(\var)  & = \{(\loc_1 \xrightarrow{\update} \loc_2,\atom,\incr) \in {\transitions \times \atoms \times \mathbb{Z}} \mid \\
    & \quad \quad \quad \quad \quad \quad \quad \quad \quad {\var^\prime \le \atom + \incr} \in \update, \atom \ne \var\}\\
    \AllCyclic(\var)  & =  \{(\loc_1 \xrightarrow{\update} \loc_2, \incr) \in {\transitions \times \mathbb{Z}} \mid {\var^\prime \le \var + \incr} \in \update, \incr > 0\}
  \end{array}
  \end{math}
 Given a path $\paath$ of $\deltaprog$ we say that $\var$ is {\em reset} on $\paath$ if there is a transition $\trans$ on $\paath$ such that $(\trans, \atom, \incr) \in \AllResets(\var)$ for some $\atom \in \atoms$ and $\incr \in \mathbb{Z}$.
\label{def:edgesets}
\end{definition}
{\em Example~B}: $\AllCyclic(k) = \{(\trans_1,1)\}$ and $\AllResets(k) = \{(\trans_0, n, 0)\}$.

I.e., we have $(\trans, \atom, \incr) \in \AllResets(\var)$ if variable $\var$ is reset to a value $\le \atom + \incr$ when executing the transition $\trans$.
Accordingly we have $(\trans, \incr) \in \AllCyclic(\var)$ if variable $\var$ is incremented by a value $\le \incr$ when executing the transition $\trans$.

Our algorithm in Definition~\ref{def:functions01} is build on a {\em mutual recursion} between the two functions $\VariableBound(\var)$ and $\TransitionBound(\trans)$, where $\VariableBound(\var)$ infers a {\em variable bound} for $\var$ and $\TransitionBound(\trans)$ infers a {\em transition bound} for the transition $\trans$.

\begin{definition}[Bound Algorithm]\label{def:functions01}
 Let $\lpathstonodedfgs: \transitions \rightarrow \vars \cup \{\onefunction\}$ be a {\em local bound mapping} for $\deltaprog$.
 We define $\VariableBound: \atoms \mapsto \expressions(\atoms)$ and
 $\TransitionBound: \transitions \mapsto \expressions(\atoms)$ as:\\
 $\VariableBound(\atom) = \atom$, if $\atom \in \atoms \setminus \vars$, else\\
  $\VariableBound(\var) = \IncrCyclic(\var) + \max\limits_{(\_, \atom, \incr) \in \AllResets(\var)} (\VariableBound(\atom) + \incr)$
  \begin{tabbing}
  $\TransitionBound(\trans) =$\=\ $\onefunction$, if $\lpathstonodedfgs(\trans) = \onefunction$, else\\
  $\TransitionBound(\trans) =$\>\ $\IncrCyclic(\lpathstonodedfgs(\trans))$\\
                              \>\ $+ \sum\limits_{(\transtwo,\atom,\incr) \in \AllResets(\lpathstonodedfgs(\trans))} \TransitionBound(\transtwo) \times \max(\VariableBound(\atom) + \incr, 0)$
 \end{tabbing}
 where\\
 $\IncrCyclic(\var) =  \sum\limits_{(\trans,\incr) \in \AllCyclic(\var)}{\TransitionBound(\trans) \times \incr}$ \ ($\IncrCyclic(\var) = 0$ for $\AllCyclic(\var) = \emptyset$)
\end{definition}

\paragraph*{Discussion}
We first explain the subroutine $\IncrCyclic(\var)$:
With $(\trans, \incr) \in \AllCyclic(\var)$ we have that a single execution of $\trans$ {\em increments} the value of $\var$ by not more than $\incr$.
$\IncrCyclic(\var)$ multiplies the transition bound of $\trans$ with the increment $\incr$ for summarizing the total amount by which $\var$ may be incremented over all executions of $\trans$.
$\IncrCyclic(\var)$ thus computes a bound on the total amount by which the value of $\var$ may be {\em incremented} during a program run.

The function $\VariableBound(\var)$ computes a variable bound for $\var$: After executing a reset transition $(\trans, \atom, \incr) \in \AllResets(\var)$, the value of $\var$ is bounded by $\VariableBound(\atom) + \incr$.
As long as $\var$ is not {\em reset}, its value cannot increase by more than $\IncrCyclic(\var)$.

The function $\TransitionBound(\trans)$ computes a transition bound for $\trans$ based on the following reasoning:
(1) The total amount by which the local bound $\lpathstonodedfgs(\trans)$ of transition $\trans$ can be {\em incremented} is bounded by $\IncrCyclic(\lpathstonodedfgs(\trans))$.
(2) We consider a reset $(\transtwo, \atom, \incr) \in \AllResets(\lpathstonodedfgs(\trans))$;
in the worst case, a single execution of $\transtwo$
resets the local bound $\lpathstonodedfgs(\transtwo)$ to $\VariableBound(\atom) + \incr$, adding $\max(\VariableBound(\atom) + \incr,0)$ to the potential number of executions of $\transtwo$;
in total all $\TransitionBound(\transtwo)$ possible executions of $\transtwo$ add up to $\TransitionBound(\transtwo) \times \max(\VariableBound(\atom) + \incr,0)$ to the potential number of executions of $\transtwo$.

\emph{Example~A}, $\lpathstonodedfgs$ as defined in Figure~\ref{fig-ex-dcps}:
$j$ is {\em reset} to $0$ on $\trans_0$ and incremented by $1$ on $\trans_1$. $i$ is reset to $n$ on $\trans_0$. Our algorithm computes $\TransitionBound(\trans_2) = \TransitionBound(\trans_1) \times 1 + \TransitionBound(\trans_0) \times 0 = \TransitionBound(\trans_1) = \TransitionBound(\trans_0) \times n = n$. Thus the overall complexity of Example~A is inferred by $\TransitionBound(\trans_1) + \TransitionBound(\trans_2) = 2n$.

\emph{Example~B}, $\lpathstonodedfgs$ as defined in Figure~\ref{fig-ex-dcps}:
$i$ and $l$ are {\em reset} to $n$ on $\trans_0$.
Our algorithm computes $\TransitionBound(\trans_1) = \TransitionBound(\trans_0) \times n = n$ and $\TransitionBound(\trans_2) = \TransitionBound(\trans_0) \times n = n$.
$j$ is {\em reset} to $0$ on $\trans_0$ and {\em reset} to $k$ on $\trans_2$.
Our algorithm computes $\TransitionBound(\trans_3) = \TransitionBound(\trans_0) \times 0 + \TransitionBound(\trans_2) \times \VariableBound(k)$.
Since $k$ is {\em reset} to $0$ on $\trans_0$ and incremented by $1$ on $\trans_1$, our algorithm computes $\VariableBound(k) = \TransitionBound(\trans_1) \times 1 = n \times 1 = n$.
Thus  $\TransitionBound(\trans_3) = \TransitionBound(\trans_2) \times \VariableBound(k) = n \times n = n^2$.
Thus the overall complexity of Example~B is inferred by $\TransitionBound(\trans_1) + \TransitionBound(\trans_2) + \TransitionBound(\trans_3) = n + n + n^2 = 2n + n^2$.

\emph{Example~2} (Figure~\ref{fig:ex1}):
$\lpathstonodedfgs = \{\trans_0,\trans_{0_a},\trans_{0_b},\trans_2 \mapsto 1, \trans_1 \mapsto y, \trans_3 \mapsto z\}$,
$\AllResets(z) = \{(\trans_2, x, 0)\}$, $\AllCyclic(x) = \{(\trans_1, 2)\}$, $\AllResets(x) = \{(\trans_{0a}, m1, 0), (\trans_{0b}, m2, 0)\}$, $\AllResets(y) = \{(\trans_0, n, 0)\}$.
We have stated the computation of $\TransitionBound(\trans_3)$ in Section~\ref{subsec:inv-and-bound-analysis}.

{\emph{Termination:} Our algorithm does not terminate if recursive calls cycle, i.e., if a call to $\TransitionBound(\trans)$ resp. $\VariableBound(\var)$ (indirectly) leads to a recursive call to $\TransitionBound(\trans)$ resp. $\VariableBound(\var)$. This can be easily detected, we return the value $\bot$ (undefined).}

\begin{theorem}[Soundness]
Let $\deltaprog(\locs, \transitions, \loc_b, \loc_e)$ be a well-defined and deterministic \dcp\ over atoms \atoms, $\lpathstonodedfgs: \transitions \mapsto {\vars \cup \{1\}}$ be a {\em local bound mapping} for $\deltaprog$, $\var \in \vars$ and $\trans \in \transitions$.
Either $\TransitionBound(\trans) = \bot$ or $\TransitionBound(\trans)$ is a {\em transition bound} for $\trans$.
Either $\VariableBound(\var) = \bot$ or $\VariableBound(\var)$ is a {\em variable bound} for $\var$.
\end{theorem}

\subsection{Context-Sensitive Bound Analysis}
\label{subsec:alg_pathsens}

So far our algorithm reasons about resets occurring on single transitions.
In this section we increase the precision of our analysis by exploiting the context under which resets are executed through a refined notion of resets and  increments.

\begin{definition}[Reset Graph]
 The {\em Reset Graph} for $\deltaprog$ is the graph $\dfg(\atoms,\edgesdfg)$ with $\edgesdfg \subseteq \atoms \times \transitions \times \mathbb{Z} \times \vars$ s.t.
 $\edgesdfg = \{(x, \trans, \incr, y) \mid {(\trans, y, \incr) \in \AllResets(x)}\}$.
 We call a {\em finite} path $\npath = \atom_n \xrightarrow{\trans_n, c_n} \atom_{n-1} \xrightarrow{\trans_{n-1}, c_{n-1}} \dots \atom_0$ in $\dfg$ with $n > 0$ a {\em reset path} of $\deltaprog$.
 We define
 $\npathin(\npath) = \atom_n$,
 $\npathincr(\npath) = \sum\limits_{i=1}^n c_i$,
 $\npathtrans(\npath) = \{\trans_n,\trans_{n-1}\dots,\trans_1\}$, and
 $\npathnodes(\npath) = \{a_n,a_{n-1}\dots,a_0\}$.
 $\npath$ is {\em sound} if for all $1 \le i < n$ it holds that $\atom_i$ is {\em reset} on all paths from the target location of $\trans_1$ to the source location of $\trans_i$ in $\deltaprog$.
 $\npath$ is {\em optimal} if $\npath$ is sound and there is no sound reset path $\hat{\npath}$ s.t. $\npath$ is a suffix of $\hat{\npath}$, i.e., $\hat{\npath} = \atom_{n+k} \xrightarrow{\trans_{n+k}, c_{n+k}} \atom_{{n+k}-1} \xrightarrow{\trans_{{n+k}-1}, c_{{n+k}-1}} \dots \atom_n \xrightarrow{\trans_n, c_n} \atom_{n-1} \xrightarrow{\trans_{n-1}, c_{n-1}} \dots \atom_0$ with $k \ge 1$.
 Let $\var \in \vars$, by $\PathSensReset(\var)$ we denote the set of optimal reset paths ending in $\var$.
 \label{def:resetgraph}
\end{definition}

\begin{figure}[t]
\begin{tabular}{c|c|c|c}
    \begin{minipage}{1.5cm}
    \center
    \begin{tikzpicture}[scale=0.4, node distance = 0.8cm, auto]
    \node (t0) {$0$};
    \node (t1) [left of=t0]{$n$};
    \node (t2) [below of=t0] {$j$};
    \node (t3) [below of=t1] {$i$};
    \path (t0) edge [line] node [right,font=\scriptsize]{$\trans_0$}  (t2)
    (t1) edge [line] node [right,font=\scriptsize]{$\trans_0$} (t3)
    ;
    \end{tikzpicture}
   \end{minipage}
 &
     \begin{minipage}{2cm}
    \center
    \begin{tikzpicture}[scale=0.4, node distance = 0.8cm, auto]
    \node (t0) {$n$};
    \node (t1b) [right of=t0, node distance=0.8cm] {$l$};
    \node (t1a) [below of=t1b, node distance=0.8cm] {$i$};
    \node (t3) [below of=t0, node distance=1.2cm] {$0$};
    \node (t1) [right of=t3, node distance=0.8cm] {$k$};
    \node (t2) [below of=t1] {$j$};

    \path (t3) edge [line] node [above,font=\scriptsize]{$\trans_0$}  (t1)
    (t0) edge [line] node [left,font=\scriptsize]{$\trans_0$}  (t1a)
    (t0) edge [line] node [above,font=\scriptsize]{$\trans_0$}  (t1b)
    (t1) edge [line] node [right,font=\scriptsize]{$\trans_2$} (t2)
    (t3) edge [line] node [above,font=\scriptsize]{$\trans_0$} (t2)
    ;
    \end{tikzpicture}
   \end{minipage}
 &
    \begin{minipage}{2cm}
   \center
    \begin{tikzpicture}[scale=0.4, node distance = 0.8cm, auto]
    \node (t00) {};
    \node (t0b) [left of=t00, node distance = 0.4cm] {$0$};
    \node (t0a)[right of=t00, node distance = 0.4cm] {$n$};
    \node (t0) [below of=t00, node distance = 0.8cm] {$r$};
    \node (t100) [right of=t0] {$i$};
    \node (t1) [below of=t0, node distance = 0.8cm] {$k$};
    \path(t0) edge [line] node [right,font=\scriptsize]{$\trans_1$} (t1)
    (t0a) edge [line] node [left,font=\scriptsize,at start]{$\trans_0$} (t0)
    (t0a) edge [line] node [right,font=\scriptsize,at start]{$\trans_0$} (t100)
    (t0b) edge [line] node [left,font=\scriptsize]{$\trans_3$} (t0)
    ;
    \end{tikzpicture}

   \end{minipage}

   &

    \begin{minipage}{2cm}
    \center
    \begin{tikzpicture}[scale=0.4, node distance = 0.8cm, auto]
    \node (t0) {$0$};
    \node (t0b) [left of=t0, node distance=0.8cm] {$0$};
    \node (t1) [below of=t0, node distance=0.8cm] {$r$};
    \node (t1b) [left of=t1] {$n$};
    \node (t3) [below of=t1, node distance = 0.8cm]  {$p$};
    \node (t3b) [below of=t1b, node distance = 0.8cm] {$x$};
    \path (t0) edge [line] node [right,font=\scriptsize]{$\trans_0$}  (t1)
    (t1) edge [line] node [right,font=\scriptsize]{$\trans_{2a}$} (t3)
    (t0b) edge [line] node [left,font=\scriptsize]{$\trans_{4}$} (t1)
    (t1b) edge [line] node [right,font=\scriptsize]{$\trans_0$}(t3b)
    ;
    \end{tikzpicture}

   \end{minipage}
\\
$\dfg(A)$ & $\dfg(B)$ & $\dfg(C)$ & $\dfg(Ex1)$\\
 \end{tabular}
 \caption{Reset Graphs, increments by $0$ are not depicted}
 \label{fig-reset-graphs}
\end{figure}
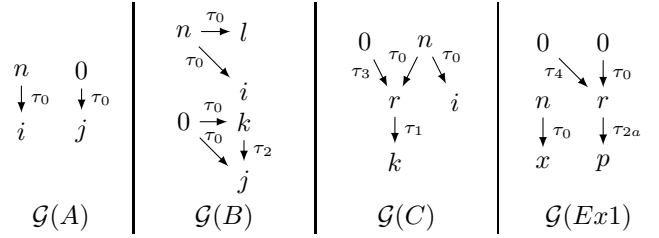

We explain the notions {\em sound} and {\em optimal} in the course of the following discussion. Figure~\ref{fig-reset-graphs} shows the reset graphs of Examples~A, B, C and Example~1 from Figure~\ref{fig:ex1}.
For a given reset $(\trans, \atom, \incr) \in \AllResets(\var)$, the reset graph determines which atom flows into variable $\var$ under which context.
For example, consider $\dfg(C)$:
When executing the reset $(\trans_1,r,0) \in \AllResets(k)$ under the context $\trans_3$, $k$ is set to $0$, if the same reset is executed under the context $\trans_0$, $k$ is set to $n$.
Note that the reset graph does not represent {\em increments} of variables. We discuss how we handle increments below.

We assume that the reset graph is a DAG.
We can always force the reset graph to be a DAG {by abstracting the \dcp: we remove all program variables which have cycles in the reset graph and all variables whose values depend on these variables.} Note that if the reset graph is a DAG, the set $\PathSensReset(\var)$ is finite for all $\var \in \vars$.

Let $\var \in \vars$. Given a reset path $\npath$ of length $k$ that ends in $\var$, we say that $(\npathtrans(\npath),\npathin(\npath),\npathincr(\npath))$ is a reset of $\var$ with context of length $k-1$.
I.e., $\AllResets(\var)$ from Definition~\ref{def:edgesets} is the set of {\em context-free} resets of $\var$ (context of length $0$), because $(\npathtrans(\npath),\npathin(\npath),\npathincr(\npath)) \in \AllResets(\var)$ iff $\npath$ ends in $\var$ and has length $1$.
Our algorithm from Definition~\ref{def:functions01} reasons {\em context free} since it uses only {\em context-free} resets.

Consider Example~C.
The precise bound for $\trans_2$ is $n$ because we can iterate $\trans_2$ only in the first iteration of the loop at $\loc_1$ since $r$ is reset to $0$ on $\trans_3$.
But when reasoning context-free, our algorithm infers a {\em quadratic} bound for $\trans_2$:
We assume $\lpathstonodedfgs$ to be given as stated in Figure~\ref{fig-ex-dcps}.
In $\dfg(C)$ $\npath = r \xrightarrow{\trans_1,0} k$ is the only reset path of length $1$ ending in $k$. Thus $\AllResets(k) = \{(\trans_1,r,0)\}$.
Our algorithm from Definition~\ref{def:functions01} computes:
$\TransitionBound(\trans_1) = \TransitionBound(\trans_0) \times n = n$, $\VariableBound(r) = \TransitionBound(\trans_0) \times n + \TransitionBound(\trans_3) \times 0 = n$, $\TransitionBound(\trans_2) = \TransitionBound(\trans_1) \times \VariableBound(r) = n \times n = n^2$.

We show how our algorithm infers the {\em linear} bound for $\trans_2$ when using {\em resets with context}:
If we consider $\npath$ with contexts, we get $\npath_1 = 0 \xrightarrow{\trans_3,0} r \xrightarrow{\trans_1,0} k$ and $\npath_2 = n \xrightarrow{\trans_0,0} r \xrightarrow{\trans_1,0} k$.
Note that $\npath_1$ and $\npath_2$ are {\em sound} by Definition~\ref{def:resetgraph} because $r$ is reset on all paths from the target location $\loc_2$ of $\trans_1$ to the source location $\loc_1$ of $\trans_1$ in Example~C (namely on $\trans_3$).
Thus $\PathSensReset(k) = \{(\{\trans_3,\trans_1\}, 0, 0), (\{\trans_0,\trans_1\}, n, 0)\}$.
We can compute a bound on the number of times that a sequence $\trans_1,\trans_2,\dots\trans_n$ of transitions may occur on a run by computing
$\min\limits_{1\le i \le n}\TransitionBound(\trans_i)$.
Thus, basing our analysis on $\PathSensReset(k)$ rather than $\AllResets(k)$ we compute:
$\TransitionBound(\trans_2) = \min(\TransitionBound(\trans_3),\TransitionBound(\trans_1)) \times 0 + \min(\TransitionBound(\trans_0),\TransitionBound(\trans_1)) \times n = \min(n, 1) \times n = n$.

We have demonstrated that our analysis gains precision when adding context to our notion of resets.
It is, however, not sound to base the analysis on maximal reset paths (i.e., resets with maximal context) only: Consider Example~B with $\lpathstonodedfgs$ as stated in Figure~\ref{fig-ex-dcps}.
There are 2 maximal reset paths ending in $j$ (see $\dfg(B)$): $\npath_1 = 0 \xrightarrow{\trans_0,0} j$ and $\npath_2 = 0 \xrightarrow{\trans_0,0} k \xrightarrow{\trans_2,0} j$.
Thus $\PathSensReset(j)^\prime = \{(\{\trans_0,\trans_2\}, 0, 0), (\{\trans_0\}, 0, 0)\}$ is the set of resets of $j$ with {\em maximal} context.
Using $\PathSensReset(j)^\prime$ rather than $\AllResets(j)$ our algorithm computes:
$\TransitionBound(\trans_3) = \min(\TransitionBound(\trans_0),\TransitionBound(\trans_2)) \times 0 + \TransitionBound(\trans_0) \times 0  + \TransitionBound(\trans_1) \times 1= \TransitionBound(\trans_1) \times 1 = n$, but $n$ is not a transition bound for $\trans_3$.
The reasoning is unsound because $\npath_2$ is {\em unsound} by Definition~\ref{def:resetgraph}:
$k$ is {\em not} reset on all paths from the target location $\loc_1$ of $\trans_2$ to the source location $\loc_1$ of $\trans_2$ in Example~B:
e.g., the path $\trans_2 = \loc_1 \xrightarrow{\update_2} \loc_1$ of Example~B does not reset $k$.

We base our {\em context sensitive} algorithm on the set $\PathSensReset(\var)$ of {\em optimal} reset paths.
The optimal reset paths are those that are maximal within the {\em sound} reset paths (Definition~\ref{def:resetgraph}).

\begin{definition}[Bound Algorithm with Context]
 Let $\lpathstonodedfgs: \transitions \rightarrow \vars \cup \{\onefunction\}$ be a {\em local bound mapping} for $\deltaprog$.
 Let $\VariableBound: \atoms \mapsto \expressions(\atoms)$ be as defined in Definition~\ref{def:functions01}.
 We override the definition of $\TransitionBound: \transitions \mapsto \expressions(\atoms)$ in Definition~\ref{def:functions01} by stating:
\begin{tabbing}
 $\TransitionBound(\trans) = $\=\ $\onefunction$  if $\lpathstonodedfgs(\trans) = \onefunction$ else\\
 $\TransitionBound(\trans) = $
                              \>\ $\sum\limits_{\npath \in \PathSensReset(\lpathstonodedfgs(\trans))}$ \= $\TransitionBound(\npathtrans(\npath)) \times \max(\VariableBound(\npathin(\npath)) + \npathincr(\npath), 0)$\\
                              \>\>$+ \sum\limits_{\atom \in \npathnodes(\npath)} \IncrCyclic(\atom)$
\end{tabbing}
 where\\
 $\TransitionBound(\{\trans_1,\trans_2,\dots,\trans_n\}) = \min\limits_{1 \le i \le n}\TransitionBound(\trans_i)$%
 \label{def:functions02}
\end{definition}

\paragraph*{Discussion and Example}
The main difference to the definition of $\TransitionBound(\trans)$ in Definition~\ref{def:functions01} is that the term $\IncrCyclic(\lpathstonodedfgs(\trans))$ is replaced by the term $\sum\limits_{\atom \in \npathnodes(\npath)} \IncrCyclic(\atom)$.
Consider the abstracted \dcp\ of Example~1 in Figure~\ref{fig:ex1}. We have discussed in Section~\ref{subsec:motiv_amortized} that $r$ may be incremented on $\trans_1$ between the reset of $r$ to $0$ on $\trans_0$ resp. $\trans_4$ and the reset of $p$ to $r$ on $\trans_{2a}$. The term $\sum\limits_{\atom \in \npathnodes(\npath)} \IncrCyclic(\atom)$ takes care of such increments which may increase the value that finally flows into $\lpathstonodedfgs(\trans)$ (in the example $p$) when the last transition on $\npath$ (in the example $\trans_{2a}$) is executed: We use the local bound mapping
$\lpathstonodedfgs = \{\trans_0 \mapsto 1,
\trans_1 \mapsto x,
\trans_{2a} \mapsto x,
\trans_{2b} \mapsto x,
\trans_4 \mapsto x,
\trans_5 \mapsto x,
\trans_3 \mapsto p\}$ for Example~1.
The reset graph of Example~1 is shown in Figure~\ref{fig-reset-graphs}. We have $\PathSensReset(p) = \{0 \xrightarrow{\trans_0} r \xrightarrow{\trans_{2a}} p, 0 \xrightarrow{\trans_4} r \xrightarrow{\trans_{2a}} p\}$.
Thus our algorithm computes $\TransitionBound(\trans_3) =
\sum\limits_{\npath \in \PathSensReset(p)} \TransitionBound(\npathtrans(\npath)) \times \max(\VariableBound(\npathin(\npath)) + \npathincr(\npath), 0)
                              + \sum\limits_{\atom \in \npathnodes(\npath)} \IncrCyclic(\atom) = \TransitionBound(\{\trans_0,\trans_{2a}\}) \times \max(\VariableBound(0),0) + \IncrCyclic(r) + \TransitionBound(\{\trans_4,\trans_{2a}\}) \times \max(\VariableBound(0),0) + \IncrCyclic(r) = 2 \times \IncrCyclic(r) = 2 \times \TransitionBound(\trans_1) \times 1 = 2 \times n$ (with $\TransitionBound(\trans_1) = n$).

\paragraph*{Complexity}
In theory there can be exponentially many resets in $\PathSensReset(\var)$. In our experiments this never occurred, enumeration of (optimal) reset paths did not affect performance.

\paragraph*{Further Optimization} We have shown in Section~\ref{sec:motivation} that transitions $\trans_3$ of Example~1 has a {\em linear} bound, precisely $n$. The Bound $2n$ that is computed by our bound algorithm from Definition~\ref{def:functions02} is {\em linear} but not precise. We compute $2n$ because $r$ appears on both reset paths of $p$ and therefore $\IncrCyclic(r) = n$ is added twice. However, there is only one transition ($\trans_{2a}$) on which $p$ is reset to $r$ and between any two executions of $\trans_{2a}$ $r$ will be reset to $0$. For this reason each increment of $r$ can only contribute once to the increase of the local bound $p$ of $\trans_3$, and not twice.
We thus suggest to further optimize our algorithm from Definition~\ref{def:functions02} by distinguishing if there is more than one way how $\atom \in \npathnodes(\npath)$ may flow into the target variable of $\npath$ or not. We divide $\npathnodes(\npath)$ into two disjoint sets
$\npathnodes_2(\npath) = \{\atom \in \npathnodes(\npath) \mid \text{more than 1 path from }  \atom \text{ to target variable of } \npath \text{ in } \dfg(\deltaprog)\}$,
$\npathnodes_1(\npath) = \npathnodes(\npath) \setminus \npathnodes_2(\npath)$. We define
\begin{tabbing}
   $\TransitionBound(\trans) = $\=\ ($\sum\limits_{\atom \in \bigcup\limits_{\npath \in \PathSensReset(\lpathstonodedfgs(\trans))}\npathnodes_1(\npath)} \IncrCyclic(\atom))~+ $\\
                                \>\ $\sum\limits_{\npath \in \PathSensReset(\lpathstonodedfgs(\trans))}$ \= $\TransitionBound(\npathtrans(\npath)) \times \max(\VariableBound(\npathin(\npath)) + \npathincr(\npath), 0)$\\
                                \>\>$+ \sum\limits_{\atom \in \npathnodes_2(\npath)} \IncrCyclic(\atom)$
\end{tabbing}
for $\lpathstonodedfgs(\trans) \ne 1$. Note that for Example~1 $\npathnodes_1(\npath) = \{r\}$ and $\npathnodes_2(\npath) = \emptyset$ for both $\npath \in \PathSensReset(p)$. Therefore $\TransitionBound(\trans_3) = \AllCyclic(r) = n$ with the optimization.

\begin{theorem}[Soundness of Bound Algorithm with Context]
Let $\deltaprog(\locs, \transitions, \loc_b,\loc_e)$ be a well-defined and deterministic \dcp\ over atoms \atoms, $\lpathstonodedfgs: \transitions \mapsto {\vars \cup \{1\}}$ be a {\em local bound mapping} for $\deltaprog$, $\var \in \vars$ and $\trans \in \transitions$.
Let $\TransitionBound(\trans)$ and $\VariableBound(\atom)$ be defined as in Definition~\ref{def:functions02}.
Either $\TransitionBound(\trans) = \bot$ or $\TransitionBound(\trans)$ is a {\em transition bound} for $\trans$.
Either $\VariableBound(\var) = \bot$ or $\VariableBound(\var)$ is a {\em variable bound} for $\var$.
\end{theorem}

\subsection{\dcp s over non-well-founded domains}
\label{subsec:alg-nonwellfounded}
In real world code, many data types are not well-founded.
The abstraction of a concrete program is much simpler and more information is kept if the abstract program model is not limited to a well-founded domain.
Below we extend our program model from Definition~\ref{def:dcp} to the non-well-founded domain $\mathbb{Z}$ by adding guards to the transitions in the program.
Interestingly our bound algorithm from Definition~\ref{def:functions01} resp. Definition~\ref{def:functions02} remains sound for the extended program model, if we adjust our notion of a {\em local transition bound} (Definition~\ref{def:potfun2}).

We extend the range of the \emph{valuations} $\vals_\atoms$ of $\atoms$ from $\mathbb{N}$ to $\mathbb{Z}$ and allow constants to be integers, i.e., we define $\atoms = \vars \cup \symbconst \cup \mathbb{Z}$.
We extend Definition~\ref{def:dcp} as follows:
The transitions $\transitions$ of a {\em guarded \dcp} $\deltaprog(\locs, \transitions,\loc_b, \loc_e)$ are a subset of $\locs \times 2^{\vars} \times 2^{\dcset(\atoms)} \times \locs$. 
A sequence  $(\loc_b,\val_0) \xrightarrow{\guard_0,\update_0} (\loc_1,\val_1) \xrightarrow{\guard_1,\update_1} \cdots$ is a {\em run} of $\deltaprog$ if it meets the conditions required in Definition~\ref{def:dcp} and additionally $\val_i(x) > 0$ holds for all ${x} \in \guard_i$.
For examples see Figure~\ref{fig:ex1}.

\begin{definition}[Local Transition Bound for \dcp s with guards]\label{def:potfun2}
Let $\deltaprog(\locs, \transitions,\loc_b, \loc_e)$ be a \dcp\ with guards over $\atoms$. Let $\trans \in \transitions$ and $\var \in \vars$. $\var$ is a {\em local bound} for $\trans$ if for all finite runs
$\trace = (\loc_b,\val_0) \xrightarrow{\trans_0} (\loc_1,\val_1) \xrightarrow{\trans_1} \cdots (\loc_e,\val_n)$ of $\deltaprog$ it holds that
$\numberOccTrans(\trans, \trace) \le \numberOccDecr(\max(\var,0), \trace)$.
\end{definition}

The algorithms in Sections~\ref{subsec:findpotentials} and~\ref{sec:abstraction} are based on the extended program model over $\mathbb{Z}$, it is straightforward to adjust them for \dcp s without guards.

\subsection{Determining Local Bounds}
\label{subsec:findpotentials}
We call a path of a \dcp\ $\deltaprog(\locs,\transitions,\loc_b,\loc_e)$ {\em simple and cyclic} if it has the same start- and end-location and does not visit a location twice except for the start- and end-location.
Given a transition $\trans \in \transitions$ we assign it $\var \in \vars$ as local bound if for all simple and cyclic paths $\paath = \loc_1 \xrightarrow{\guard_1, \update_1} \loc_2 \xrightarrow{\guard_2, \update_2}...\loc_n$ ($\loc_n = \loc_1$) of \deltaprog\  that  traverse $\trans$ it holds that
(1) $\exists 0 < i < n$ s.t.  $\var \in \guard_i$ and
(2) $\exists 0 < i < n$ s.t.  $\var^\prime \le \var + \incr \in \update_i$ for some $\incr < 0$.
Our implementation avoids an explicit enumeration of the simple and cyclic paths of $\deltaprog$ {{by a simple data flow analysis}.

\section{Program Abstraction}
\label{sec:abstraction}

In this section we present our concrete program model and discuss how we abstract a given program to a \dcp.

\begin{definition}[Program]
\label{def:program}
  Let $\states$ be a set of \emph{states}.
  The set of \emph{transition relations} $\trns = 2^{\states \times \states}$ is the set of relations over $\states$.
  A \emph{program} is a directed labeled graph $\prog = (\locs, \edges, \loc_b,\loc_e)$, where $\locs$ is a finite set of \emph{locations}, $\loc_b \in \locs$ is the entry location, $\loc_e \in \locs$ is the exit location and $\edges \subseteq \locs \times \trns \times \locs$ is a finite set of \emph{transitions}.
  We write $\loc_1 \xrightarrow{\trn} \loc_2$ to denote a transition $(\loc_1,\trn,\loc_2)$.
  
  A \emph{norm} $\norm \in \states \rightarrow \mathbb{Z}$ is a function that maps the states to the integers.
\end{definition}

Programs are labeled transition systems over some set of states, where each transition is labeled by a transition relation that describes how the state changes along the transition.
Note, that a \dcp\ (Definition~\ref{def:dcp}) is a program by Definition~\ref{def:program}.

%

\begin{definition}[Transition Invariants]
  Let $\norm_1,\norm_2,\norm_3 \in \states \rightarrow \mathbb{Z}$ be norms, and let $c \in \mathbb{Z}$ be some integer.
  We say $\norm_1^\prime \le \norm_2 + \norm_3$  is \emph{invariant for} $\loc_1 \xrightarrow{\trn} \loc_2$, if $\norm_1(\state_2) \le \norm_2(\state_1) + \norm_3(\state_1)$ holds for all $(\state_1, \state_2) \in \rho$. 
  We say $\norm_1 > 0$ is \emph{invariant for} $\loc_1 \xrightarrow{\trn} \loc_2$, if $\norm_1(\state_1) > 0$ holds for all $(\state_1, \state_2) \in \rho$.
\end{definition}

\begin{definition}[Abstraction of a Program]
Let $\prog = (\locs, \edges, \loc_b, \loc_e)$ be a program and let $\norms$ be a finite set of norms.
A \dcp\ $\deltaprog = (\locs, \edgesP, \loc_b, \loc_e)$ with atoms $\norms$ is an abstraction of the program $\prog$ iff for each transition ${\loc_1 \xrightarrow{\trn} \loc_2} \in \edges$ there is a transition ${\loc_1 \xrightarrow{\update,\guard} \loc_2} \in \edgesP$
s.t. every ${\norm_1^\prime \le \norm_2 + c} \in \update$ is invariant for ${\loc_1 \xrightarrow{\trn} \loc_2}$ and for every $\norm_1 \in \guard$ it holds that $\norm_1 > 0$ is invariant for ${\loc_1 \xrightarrow{\trn} \loc_2}$.
\end{definition}

We propose to abstract a program $\prog = (\locs, \edges,\loc_b,\loc_e)$ to a \dcp\  $\deltaprog = (\locs, \edgesP,\loc_b,\loc_e)$ as follows:
Let $\norms$ be some initial set of norms.\\
1) For each transition ${\loc_1 \xrightarrow{\trn} \loc_2} \in \edges$ we generate a set of difference constraints $\alpha(\trn)$: Initially we set $\alpha(\trn) = \emptyset$ for all transitions ${\loc_1 \xrightarrow{\trn} \loc_2}$.
    We then repeat the following construction until the set of norms $\norms$ becomes stable:
    For each $\norm_1 \in \norms$ and ${{\loc_1 \xrightarrow{\trn} \loc_2}} \in \edges$ we check whether there is a difference constraint of form $\norm_1^\prime \le \norm_2 + \incr$ for $\norm_1$ in $\alpha(\trn)$.
    If not, we try to find a norm $\norm_2$ (possibly not yet in $\norms$) and a constant $\incr \in \mathbb{Z}$ s.t. $\norm_1^\prime \le \norm_2 + \incr$ is invariant for $\trn$.
    If we find appropriate $\norm_2$ and $\incr$, we add $\norm_1^\prime \le \norm_2 + \incr$ to $\alpha(\trn)$ and $\norm_2$ to $\norms$. I.e., our transition abstraction algorithm performs a fixed point computation which might not terminate if new terms keep being added (see discussion in next section).\\
2) For each transition ${\loc_1 \xrightarrow{\trn} \loc_2}$ we generate a set of guards $\guards(\trn)$: Initially we set $\guards(\trn) = \emptyset$ for all transitions ${\loc_1 \xrightarrow{\trn} \loc_2}$.
   For each $\norm \in \norms$ and each transition ${\loc_1 \xrightarrow{\trn} \loc_2}$ we check if $\norm > 0$ is invariant for ${\loc_1 \xrightarrow{\trn} \loc_2}$. If so, we add $\norm$ to $\guards(\trn)$.\\
3) We set $\edgesP = \{{\loc_1 \xrightarrow{\guards(\trn),\alpha(\trn)} \loc_2} \mid {\loc_1 \xrightarrow{\trn} \loc_2} \in \edges\}$.

In the following we discuss how we implement the above sketched abstraction algorithm.

\subsection{Implementation}

\paragraph*{0. Guessing the initial set of Norms.}
We aim at creating a suitable abstract program for bound analysis. In our non-recursive setting, complexity evolves from iterating loops.
Therefore we search for expressions which limit the number of loop iterations. For this purpose
we consider conditions of form $a > b$ resp. $a \ge b$ found in loop headers or on loop-paths if they involve loop counter variables, i.e., variables which are incremented and/or decremented inside the loop. Such conditions are likely to limit the
consecutive execution of single or multiple loop-paths. From each such condition we form the integer expression $b - a$ and add it to our initial set of norms. Note that on those transitions on which $a > b$ holds, $b - a > 0$ must hold.

\paragraph*{1. Abstracting Transitions.}
For a given norm $\norm \in \norms$ and a transition ${\loc_1 \xrightarrow{\trn} \loc_2}$ we derive a transition predicate ${\norm^\prime \le \norm_2 + \incr} \in \alpha(\trn)$ as follows:
We symbolically execute $\trn$ for deriving $\norm^\prime$ from $\norm$.
In order to keep the number of norms low, we first try\\
i) to find a norm $\norm_2 \in \norms$ s.t. $\norm^\prime \le \norm_2 + \norm_3$ is invariant for $\rho$ where $\norm_3$ is some integer valued expression.
If $\norm_3 = \incr$ for some integer $\incr \in \mathbb{Z}$ we derive the transition predicate $\norm^\prime \le \norm_2 + \incr$. Else
we use our bound algorithm (Section~\ref{sec-alg}) for over-approximating $\norm_3$ by a constant expression $k \ge \norm_3$ and infer the transition predicate $\norm^\prime \le \norm_2 + k$ where we consider $k$ to be a symbolic constant.\\
ii) If i) fails, we form a norm $\norm_4$ s.t. $\norm^\prime \le \norm_4 + \incr$ by separating constant parts in the expression $\norm^\prime$ using associativity and commutativity of the addition operator.
E.g., given $\norm^\prime = \var + 5$ we set $\norm_4 = \var$ and $\incr = 5$.
We add $\norm_4$ to $\norms$ and derive the predicate $\norm^\prime \le \norm_4 + \incr$.

Since case ii) triggers a recursive abstraction for the newly added norm we have to ensure the termination of our abstraction procedure: Note that we can always stop the abstraction process at any point, getting a sound abstraction of the original program.
We therefore enforce termination of the abstraction algorithm by limiting the chain of recursive abstraction steps triggered by entering case ii) above: In case this limit is exceeded we remove all norms from the abstract program which
form part of the limit exceeding chain of recursive abstraction steps. This also ensures well-definedness of the resulting abstract program.

Further note that the \dcp s generated by our algorithm are always {\em deterministic}: For each transition, we get at most one predicate ${\norm^\prime \le \norm_2 + \incr}$ for each $\norm \in \norms$.

\paragraph*{2. Inferring Guards}
Given a transition $\loc_1 \xrightarrow{\rho} \loc_2$ and a norm $\norm$, we use an SMT solver to check whether $\norm > 0$ is invariant for $\loc_1 \xrightarrow{\rho} \loc_2$. If so, we add $\norm$ to $\guards(\rho)$.

\paragraph*{Non-linear Iterations.} We handle counter updates such as $x^\prime = 2x$ or $x^\prime = x/2$ {as discussed in~\cite{sinn2014simple}}.

\section{Experiments} \label{sec:experiments}

\begin{figure}{\scriptsize
\begin{tabular}{|l|l|l|l|l|l|l|l|l|l|}
\hline
		& \textbf{Succ.} 						& $1$ 		&  $n$ 	& $n^2$	& $n^3$	&  $n^{> 3}$ 	& $2^n$					& Time						& TO   \\\hline
Loopus'15	& 806							& 205						& 489		 			 &		97	 		&	13	 			&	2	   				&0	 				& 15m 						& 6   \\\hline
Loopus'14	& 431							& 200						& 188		 			 &		43	 		&	0	 			&	0	   				&0	 				& 40m 						& 20   \\\hline
KoAT		& 430							& 253						& 138		 			 &		35	 		&	2	 			&	0	   				&2 	 				& 5,6h 						& 161  \\\hline
CoFloCo		& 386							& 200						& 148		 			 & 		38	 		&	0	 			&	0	   				&0	 				& 4.7h 						& 217 \\\hline
\end{tabular}}
\caption{Tool Results on analyzing the complexity of 1659 functions in the cBench benchmark, none of the tools infers $\mathit{log}$ bounds.}
\vspace*{-4mm}
\label{ComparisonResults}
\end{figure}

\paragraph*{Implementation}
We have implemented the presented algorithm into our tool Loopus~\cite{loopuswebsite}.
Loopus reads in the LLVM~\cite{llvm} intermediate representation and performs an intra-procedural analysis.
It is capable of computing bounds for loops as well as analyzing the complexity of non-recursive functions.

\paragraph*{Experimental Setup}
For our experimental comparison we used the program and compiler optimization benchmark {\em Collective Benchmark}~\cite{cbench} (cBench), which contains a total of 1027 different C files (after removing code duplicates) with 211.892 lines of code.
In contrast to our earlier work we did not perform a loop bound analysis but a complexity analysis on function level.
We set up the first comparison of complexity analysis tools on real world code.
For comparing our new tool (Loopus'15) we chose the 3 most promising tools from recent publications: the tool KoAT implementing the approach of~\cite{conf/tacas/BrockschmidtEFFG14}, the tool CoFloCo implementing~\cite{conf/aplas/Flores-MontoyaH14} and our own earlier implementation (Loopus'14) \cite{sinn2014simple}.
Note that we compared against the most recent versions of KoAT and CoFloCo {(download 01/23/15)}.\footnote{https://github.com/s-falke/kittel-koat, https://github.com/aeflores/CoFloCo}
The experiments were performed on a Linux system with an Intel dual-core 3.2 GHz processor and 16 GB memory.
We used the following experimental set up:\\
1) We compiled all 1027 C files in the benchmark into the llvm intermediate representation using clang.\\
2) We extracted all 1751 functions which contain at least one loop using the tool llvm-extract (comes with the llvm tool suite).
Extracting the functions to single files guarantees an intra-procedural setting for all tools.\\
3) We used the tool llvm2kittel~\cite{llvm2kittel} to translate the 1751 llvm modules into 1751 text files in the Integer Transition System (ITS) format read in by KoAT.\\
4) We used the transformation described in~\cite{conf/aplas/Flores-MontoyaH14} to translate the ITS format of KoAT into the ITS format of CoFloCo.
This last step is necessary because there exists no direct way of translating C or the llvm intermediate representation into the CoFloCo input format.\\
5) We decided to exclude the 91 recursive functions in the set because we were not able to run CoFloCo on these examples (the transformation tool does not support recursion), KoAT was not successful on any of them and Loopus does not support recursion.

In total our example set thus comprises 1659 functions.

\paragraph*{Evaluation}
Table~\ref{ComparisonResults} shows the results of the 4 tools on our benchmark using a time out of 60 seconds. The first column shows the number of functions which were successfully bounded by the respective tool,
the last column shows the number of time outs, on the remaining examples (not shown in the table) the respective tool did not time out but was also not able compute a bound. The column {\em Time} shows the total time used by the tool to process the benchmark.
Loopus'15 computes the complexity for about twice as many functions as KoAT, CoFloCo and Loopus'14 while needing an order of magnitude less time than KoAT and CoFloCo and significantly less time than Loopus'14.
We conclude that our approach is both scalable and more successful than existing approaches.

\paragraph*{Pointer and Shape Analysis}
Even Loopus'15, computed bounds for only about half of the functions in the benchmark. Studying the benchmark code we concluded that for many functions pointer alias and/or shape analysis is needed for inferring functional complexity.
In our experimental comparison such information was not available to the tools.
Using optimistic (but unsound) assumptions on pointer aliasing and heap layout, our tool Loopus'15 was able to compute the complexity for in total 1185 out of the 1659 functions in the benchmark (using 28 minutes total time).

\paragraph*{Amortized Complexity}
During our experiments, we found 15 examples with an amortized complexity that could only be inferred by the approach presented in this paper.
These examples and further experimental results can be found on~\cite{loopuswebsite} where our new tool is offered for download.

\FloatBarrier

\bibliographystyle{abbrv}
\bibliography{main}

\appendix

\subsection{Full Example}
\label{app-example}

\begin{figure*}
\begin{tabular}{l|l|l}

\begin{minipage}{4.7cm}
 \begin{alltt}
xnu(int len) \{
  int beg,end,i = 0;
\(l\sb{1}\) while(i < len) \{
    i++;
\(l\sb{2}\)   if (*)
      end = i;
\(l\sb{3}\)   if (*) \{
      int k = beg;
\(l\sb{4}\)     while (k < end)
        k++;
      end = i;
      beg = end;
    \}
\(l\sb{5}\) \}
\}
 \end{alltt}
\end{minipage}

&

 \begin{minipage}{5.5cm}\hspace{-0.5cm}
 \begin{tikzpicture}[scale=0.6, node distance = 2cm, auto]

\node (t0)  {$\mathit{begin}$};
\node (t1) [below of=t0, node distance = 1.5cm] {$\loc_1$};
\node (t2) [below of=t1, node distance = 1.5cm]  {$\loc_2$};
\node (t3) [below of=t2, node distance = 1.5cm] {$\loc_3$};
\node (t4) [below of=t3, right of=t3, node distance = 2cm]  {$\loc_4$};
\node (t5) [below of=t3, left of=t3, node distance = 2cm]  {$\loc_5$};
\node (t9) [left of=t1, node distance = 1.5cm]  {$\mathit{end}$};

\path
(t0) edge [line] node [right,font=\scriptsize ] {
$\trn_0 \equiv$
 \begin{tabular}{l}
 $b^\prime = 0$\\
 $e^\prime = 0$\\
 $i^\prime = 0$
 \end{tabular}
}(t1)
(t1) edge [line] node [right,font=\scriptsize ] {
$\trn_1 \equiv$
\begin{tabular}{l}
$i < l $\\
$b^\prime = b$\\
$e^\prime = e$\\
$i^\prime = i + 1$
\end{tabular}
}(t2)
(t2) edge [line] node [left,font=\scriptsize ] {
\begin{tabular}{l}
$\trn_{2b} \equiv$\\
$b^\prime = b$\\
$e^\prime = e$\\
$i^\prime = i$\\
\end{tabular}
}(t3)
(t2) edge [line, out=-30, in=30] node [right,font=\scriptsize ] {
$\trn_{2a} \equiv$
\begin{tabular}{c}
$b^\prime = b$\\
$e^\prime = i$\\
$i^\prime = i$
\end{tabular}
}(t3)
(t3) edge [line] node [right,near start,font=\scriptsize ] {
$\trn_{3a} \equiv$
\begin{tabular}{l}
$k^\prime = b$\\
$b^\prime = b$\\
$e^\prime = e$\\
$i^\prime = i$\\
\end{tabular}
} (t4)
(t3) edge [line] node [left, near start, font=\scriptsize,yshift=+1mm ] {
\begin{tabular}{l}
$\trn_{3b} \equiv$\\
$b^\prime = b$\\
$e^\prime = e$\\
$i^\prime = i$\\
\end{tabular}
}(t5)
(t4) edge [line] node [below,font=\scriptsize,xshift=-2mm] {
$\trn_5 \equiv$
\begin{tabular}{l}
$k \ge e$\\
$e^\prime = i$\\
$b^\prime = i$\\
$i^\prime = i$
\end{tabular}
}(t5)
(t5) edge [line, bend left] node [left,font=\scriptsize] {
\begin{tabular}{l}
$\trn_6 \equiv$\\
$b^\prime = b$\\
$e^\prime = e$\\
$i^\prime = i$\\
\end{tabular}
}(t1)
(t4) edge [loop below] node [below,font=\scriptsize,xshift=-1mm] {
$\trn_4 \equiv$
\begin{tabular}{l}
$k < e$\\
$k^\prime = k + 1$\\
$b^\prime = b$\\
$e^\prime = e$\\
$i^\prime = i$\\
\end{tabular}}(t4)

(t1) edge [line] node [above,font=\scriptsize] {$i \ge l$} (t9);
\end{tikzpicture}
 \end{minipage}
  &\
 \begin{minipage}{5cm}\hspace{-1cm}
 \begin{tikzpicture}[scale=0.6, node distance = 2cm, auto]

\node (t0)  {$\loc_0$};
\node (t1) [below of=t0, node distance = 1.5cm] {$\loc_1$};
\node (t2) [below of=t1, node distance = 1.5cm]  {$\loc_2$};
\node (t3) [below of=t2, node distance = 1.5cm] {$\loc_3$};
\node (t4) [below of=t3, right of=t3, node distance = 2cm]  {$\loc_4$};
\node (t5) [below of=t3, left of=t3, node distance = 2cm]  {$\loc_5$};

\path
(t0) edge [line] node [right,font=\scriptsize ] {
 \begin{tabular}{l}
 $(e - b)^\prime \le 0$;\\
 $(i - b)^\prime \le 0$;\\
 $(l - i)^\prime \le l$;
 \end{tabular}
}(t1)
(t1) edge [line] node [right,font=\scriptsize ] {
\begin{tabular}{l}
$(l - i) > 0$\\
$(e - b)^\prime \le (e - b)$\\
$(i - b)^\prime \le (i - b) + 1$\\
$(l - i)^\prime \le (l - i) - 1$\\
\end{tabular}
}(t2)
(t2) edge [line] node [left,font=\scriptsize,xshift=2mm] {
\begin{tabular}{r}
$(e - b)^\prime \le (e - b)$\\
$(i - b)^\prime \le (i - b)$\\
$(l - i)^\prime \le (l - i)$\\
\end{tabular}
}(t3)
(t2) edge [line, out=-30, in=30] node [right,font=\scriptsize ] {
\begin{tabular}{l}
$(e - b)^\prime \le (i - b)$\\
$(i - b)^\prime \le (i - b)$\\
$(l - i)^\prime \le (l - i)$\\
\end{tabular}
}(t3)
(t3) edge [line] node [right,font=\scriptsize] {
\begin{tabular}{r}
$(e-k)^\prime \le (e-b)$\\
$(e - b)^\prime \le (e - b)$\\
$(i - b)^\prime \le (i - b)$\\
$(l - i)^\prime \le (l - i)$\\
\end{tabular}} (t4)
(t3) edge [line] node [left,font=\scriptsize,at start] {
$(e - b)^\prime \le (e - b)$
}
node [left,font=\scriptsize,yshift=-3mm,at start,xshift=-2mm] {
$(i - b)^\prime \le (i - b)$
}
node [left,font=\scriptsize,yshift=-6mm,xshift=-5mm,at start] {
$(l - i)^\prime \le (l - i)$
} 
(t5)
(t4) edge [line] node [below,font=\scriptsize ] {
\begin{tabular}{l}
$(e - b)^\prime \le 0$\\
$(i - b)^\prime \le 0$\\
$(l - i)^\prime \le (l - i)$\\
\end{tabular}
}(t5)
(t5) edge [line, bend left=80] node [left,font=\scriptsize, at end,yshift=5mm] {
\begin{tabular}{r}
$(e - b)^\prime \le (e - b)$\\
$(i - b)^\prime \le (i - b)$\\
$(l - i)^\prime \le (l - i)$\\
\end{tabular}}(t1)
(t4) edge [loop below] node [below,font=\scriptsize] {
\begin{tabular}{r}
$(e - k) > 0$\\
$(e - k)^\prime \le (e - k) - 1$\\
\end{tabular}}(t4);

\end{tikzpicture}
 \end{minipage}\\
 (a) Example~3 & (b) LTS of Example~3 & \parbox{4.1cm}{(c) Abstracted \dcp\ for Example~3}\\
 \end{tabular}
 \caption{Example~3 shows the code after which we have modeled Example~1, * denotes non-determinism (arising from conditions not modeled in the analysis)}
\label{fig-ex1-lts}
\end{figure*}

Example~3 in Figure~\ref{fig-ex1-lts} contains a snippet of the source code after which we have modeled Example~1 in Figure~\ref{fig:ex1}.
Example~3 can be found in the SPEC CPU2006 benchmark\footnote{https://www.spec.org/cpu2006/}, in function XNU of 456.hmmer/src/masks.c.
The outer loop in Example~3 partitions the interval $[0,len]$ into disjoint sub-intervals $[beg, end]$.
The inner loop iterates over the sub-intervals.
Therefore the inner loop has an overall linear iteration count.
Example~3 is a natural example for amortized complexity: Though a single visit to the inner loop can cost $len$ (if $beg = 0$ and $end = len$), several visits can also not cost more than $len$ since in each visit the loop iterates over a disjoint sub-interval.
I.e., the total cost $len$ of the inner loop is the {\em amortized cost} over all visits to the inner loop.
To the best of our knowledge our new implementation Loopus'15 (available at~\cite{loopuswebsite}) is the only tool that infers the linear complexity of Example~3 without user interaction.\\

\subsubsection{Abstraction}
In Figure~\ref{fig-ex1-lts}~(b) the labeled transition system for Example~3 is shown.
We discuss how our abstraction algorithm from Section~\ref{sec:abstraction} abstracts the example to the \dcp\ shown in Figure~\ref{fig-ex1-lts}~(c).

Our heuristics add the expressions $l - i$ and $e - k$ generated from the conditions $k < e$ and $i < l$ to the initial set of norms $N$.
Thus our initial set of norms is $\norms = \{l - i, e - k\}$.

\begin{itemize}
\item We check how $l -i$ changes on the transitions $\trn_0,\trn_1,\trn_{2a},\trn_{2b},\trn_{3a},\trn_{3b},\trn_4,\trn_5,\trn_6$:
\begin{itemize}
 \item $\trn_0$: we derive $(l - i)^\prime \le l$ (reset), we add $l$ to $N$
  \item $\trn_1$: we derive $(l - i)^\prime \le (l -i) - 1$ (negative increment)
 \item $\trn_{2a},\trn_{2b},\trn_{3a},\trn_{3b},\trn_4,\trn_5,\trn_6$: $l - i$ unchanged
\end{itemize}
\item We check how $l$ changes on the transitions $\trn_0,\trn_1,\trn_{2a},\trn_{2b},\trn_{3a},\trn_{3b},\trn_4,\trn_5,\trn_6$:
\begin{itemize}
 \item unchanged on all transitions
\end{itemize}
\item We check how $e - k$ changes on the transitions $\trn_{3a},\trn_4$ ($k$ is only defined at $\loc_4$):
\begin{itemize}
\item $\trn_{3a}$: we derive $(e -k)^\prime \le (e - b)$ (reset), we add $(e - b)$ to $N$
\item $\trn_4$: we derive $(e -k)^\prime \le (e - k) - 1$ (negative increment)
\end{itemize}
\item We check how $e - b$ changes on the transitions $\trn_0,\trn_1,\trn_{2a},\trn_{2b},\trn_{3a},\trn_{3b},\trn_4,\trn_5,\trn_6$::
\begin{itemize}
\item $\trn_0$: we derive $(e - b)^\prime \le 0$ (reset)
\item $\trn_{2a}$: we derive $(e - b)^\prime \le (i - b)$, we add $(i - b)$ to $N$
\item $\trn_5$: we derive $(e - b)^\prime \le 0$ (reset)
\item $\trn_1,\trn_{2b},\trn_{3a},\trn_{3b},\trn_4,\trn_6$:: $e - b$ unchanged
\end{itemize}
\item We check how $i - b$ changes on the transitions $\trn_0,\trn_1,\trn_{2a},\trn_{2b},\trn_{3a},\trn_{3b},\trn_4,\trn_5,\trn_6$:
\begin{itemize}
\item $\trn_0$: we derive $(i - b)^\prime \le 0$ (reset)
\item $\trn_1$: we derive $(i - b)^\prime \le (i - b) + 1$ (increment)
\item $\trn_5$: we derive $(i - b)^\prime \le 0$ (reset)
\item $\trn_{2a},\trn_{2b},\trn_{3a},\trn_{3b},\trn_4,\trn_6$:: unchanged
\end{itemize}
\item We have processed all norms in $N$
\end{itemize}

We infer that $\trn_1 \models (l - i) > 0$ and $\trn_4 \models (e - k) > 0$.

The resulting \dcp\ is shown in Figure~\ref{fig-ex1-lts}(c).\\

\subsubsection{Bound Computation}
\begin{figure*}
\begin{tabular}{l|ll}
  \begin{minipage}{7cm}
 \begin{tikzpicture}[scale=0.6, node distance = 2cm, auto]

\node (t0)  {$\loc_0$};
\node (t1) [below of=t0, node distance = 1.5cm] {$\loc_1$};
\node (t2) [below of=t1, node distance = 1.5cm]  {$\loc_2$};
\node (t3) [below of=t2, node distance = 1.5cm] {$\loc_3$};
\node (t4) [below of=t3, right of=t3, node distance = 2cm]  {$\loc_4$};
\node (t5) [below of=t3, left of=t3, node distance = 2cm]  {$\loc_5$};

\path
(t0) edge [line] node [right,font=\scriptsize ] {
 $\trans_0 \equiv $\begin{tabular}{l}
 $q^\prime \le 0$;\\
 $r^\prime \le 0$;\\
 $x^\prime \le l$;
 \end{tabular}
}(t1)
(t1) edge [line] node [right,font=\scriptsize ] {
$\trans_1 \equiv$ \begin{tabular}{l}
$x > 0$\\
$q^\prime \le q$\\
$r^\prime \le r + 1$\\
$x^\prime \le x - 1$\\
\end{tabular}
}(t2)
(t2) edge [line] node [left,font=\scriptsize ] {
\begin{tabular}{c}
$\trans_{2b}\equiv$\\
$q^\prime \le q$\\
$r^\prime \le r$\\
$x^\prime \le x$\\
\end{tabular}
}(t3)
(t2) edge [line, out=-30, in=30] node [right,font=\scriptsize ] {
$\trans_{2a} \equiv $\begin{tabular}{l}
$q^\prime \le r$\\
$r^\prime \le r$\\
$x^\prime \le x$\\
\end{tabular}
}(t3)
(t3) edge [line] node [right,near start,font=\scriptsize ] {
$\trans_{3a} \equiv $\begin{tabular}{l}
$p^\prime \le q$\\
$q^\prime \le q$\\
$r^\prime \le r$\\
$x^\prime \le x$\\
\end{tabular}} (t4)
(t3) edge [line] node [near end,right,font=\scriptsize ] {
$\trans_{3b}\equiv$
\begin{tabular}{c}
$q^\prime \le q$\\
$r^\prime \le r$\\
$x^\prime \le x$\\
\end{tabular}
}(t5)
(t4) edge [line] node [below,font=\scriptsize ] {
$\trans_5 \equiv $\begin{tabular}{c}
$q^\prime \le 0$\\
$r^\prime \le 0$\\
$x^\prime \le x$\\
\end{tabular}
}(t5)
(t5) edge [line, bend left=70] node [left,near end,xshift=-1mm,font=\scriptsize ] {
$\trans_6 \equiv$\begin{tabular}{r}
$q^\prime \le q$\\
$r^\prime \le r$\\
$x^\prime \le x$\\
\end{tabular}}(t1)
(t4) edge [loop below] node [below,font=\scriptsize] {
$\trans_4 \equiv$
\begin{tabular}{l}
$p > 0$\\
$p^\prime \le p - 1$\\
\end{tabular}}(t4);

\end{tikzpicture}
 \end{minipage}
 & 
 \ \ 
 \begin{minipage}{1.2cm}
   \begin{tikzpicture}[scale=0.6, node distance = 1cm, auto]
    \node (t0)  {$\mathit{l}$};
    \node (t1) [below of=t0] {$x$};
    \path(t0) edge [line] node [right,font=\scriptsize ] {$\trans_0$}(t1)
    ;
   \end{tikzpicture}
 \end{minipage}

 &
 \begin{minipage}{4.5cm}
 \begin{tikzpicture}[scale=0.4, node distance = 1cm, auto]
    \node (t1) [] {$0$};
    \node (t2) [left of=t1] {$0$};
    \node (t3) [below of=t1] {$r$};
    \node (t4) [left of=t3] {$0$};
    \node (t5) [left of=t4] {$0$};
    \node (t6) [below of=t4] {$q$};
    \node (t7) [below of=t6] {$p$};
    \path(t1) edge [line] node [right,font=\scriptsize ] {$\trans_0$}(t3)
    (t2) edge [line] node [right,font=\scriptsize ] {$\trans_5$}(t3)
    (t3) edge [line] node [right,font=\scriptsize ] {$\trans_{2a}$}(t6)
    (t4) edge [line] node [right,near start,font=\scriptsize ] {$\trans_5$}(t6)
    (t5) edge [line] node [left,font=\scriptsize ] {$\trans_0$}(t6)
    (t6) edge [line] node [right,font=\scriptsize ] {$\trans_{3a}$}(t7)
    ;
   \end{tikzpicture}
 \end{minipage}
 \\
 
 & &\\
	 \dcp\ for Example~3, variables renamed  & & \ \ Reset Graph
 \end{tabular}
 \caption{}
 \label{fig-ex-dcp-ex-c}
\end{figure*}

We discuss how our bound algorithm from Section~\ref{sec-alg} infers the {\em linear} bound for the inner loop at $\loc_4$.
For ease of readability, we state the abstracted \dcp\ of Example~3 in Figure~\ref{fig-ex-dcp-ex-c} renaming the variables by the following scheme: $\{\mathbf{p} = (e - k), \mathbf{q} = (e - b), \mathbf{r} = (i - b), \mathbf{x} = (l - i)\}$.
On the right hand side the reset graph is shown.
Our Algorithm from Definition~\ref{def:functions02} now computes a bound for the example by the following reasoning:	
\begin{enumerate}
\item Our algorithm for determining the local bound mapping (Section~\ref{subsec:findpotentials})
    assigns the following local bounds to the respective transitions $\lpathstonodedfgs(\trans_0) = 1$, $\lpathstonodedfgs(\trans_1) = \lpathstonodedfgs(\trans_{2a}) = \lpathstonodedfgs(\trans_{2b}) = \lpathstonodedfgs(\trans_{3a}) = \lpathstonodedfgs(\trans_{3b}) = \lpathstonodedfgs(\trans_5) = \lpathstonodedfgs(\trans_6) = x$,
$\lpathstonodedfgs(\trans_4) = p$.
\item $\PathSensReset(p) = \{0 \xrightarrow{\trans_0,0} r \xrightarrow{\trans_{2a}, 0} q \xrightarrow{\trans_{3a},0} p, 0 \xrightarrow{\trans_5,0} r \xrightarrow{\trans_{2a}, 0} q \xrightarrow{\trans_{3a},0} p,
0 \xrightarrow{\trans_0,0} q \xrightarrow{\trans_{3a},0} p, 0 \xrightarrow{\trans_5,0} q \xrightarrow{\trans_{3a},0} p\}$
\item We get: $\TransitionBound(\trans_1)$ resp. $\TransitionBound(\trans_{2a})$ resp. $\TransitionBound(\trans_{2b})$ resp. $\TransitionBound(\trans_{3a})$ resp. $\TransitionBound(\trans_{3b})$ resp. $\TransitionBound(\trans_5)$ resp. $\TransitionBound(\trans_6)= \TransitionBound(\trans_0) \times l = l$ (Definition~\ref{def:functions02}) with $\TransitionBound(\trans_0) = 1$
\item For $\trans_4$ we get: $\TransitionBound(\trans_4) = \TransitionBound(\trans_0, \trans_{2a}, \trans_{3a}) \times 0 + \TransitionBound(\trans_1) \times 1 + \TransitionBound(\trans_5, \trans_{2a}, \trans_{3a}) \times 0 + \TransitionBound(\trans_1) \times 1 + \TransitionBound(\trans_0, \trans_{3a}) \times 0 + \TransitionBound(\trans_5, \trans_{3a}) \times 0 =
n \times 1 + n \times 1 = 2n$ (Definition~\ref{def:functions02}) with $\TransitionBound(\trans_1) = n$
\item We get the precise bound $n$ for $\trans_4$ when applying the optimization presented in the discussion under Definition~\ref{def:functions02}: For all $\npath \in \PathSensReset(p)$ we have $\npathnodes_1(\npath) = \{r,q\}$ and $\npathnodes_2(\npath) = \emptyset$. Therefore $\TransitionBound(\trans_4) = \TransitionBound(\trans_1) \times 1 + \TransitionBound(\trans_0, \trans_{2a}, \trans_{3a}) \times 0 + \TransitionBound(\trans_5, \trans_{2a}, \trans_{3a}) \times 0 + \TransitionBound(\trans_0, \trans_{3a}) \times 0 + \TransitionBound(\trans_5, \trans_{3a}) \times 0 = n \times 1 =
n$ with $\TransitionBound(\trans_1) = n$.
\end{enumerate}

\end{document}